\documentclass[11pt,a4paper]{article}

\usepackage{enumitem}
 \usepackage{comment}
\usepackage[left=3cm,top=4cm,bottom=2cm,right=3cm]{geometry}
\usepackage[colorlinks=true,urlcolor = purple,linkcolor=teal,citecolor=magenta,bookmarks=true]{hyperref}
\usepackage{inputenc}
\usepackage{cite}

\usepackage{empheq} 
\usepackage{bm}
\usepackage{amsmath}
\usepackage{amsfonts,amsthm,amssymb}
\usepackage{mathtools}
\usepackage{esint}
\usepackage{graphicx}
\usepackage{xcolor}

\newcommand{\red}[1]{{ #1}}

\newcommand{\bc}{\begin{center}}
\newcommand{\ec}{\end{center}}
\def\ba#1{\begin{array}{#1}\displaystyle}
\newcommand{\ea}{\end{array}}

\newcommand{\beq}{\begin{equation}}
\newcommand{\eeq}{\end{equation}}
\newcommand{\beqa}{\begin{eqnarray}}
\newcommand{\eeqa}{\end{eqnarray}}
\newcommand{\no}{\nonumber}
\newcommand{\n}{\nonumber\\}
\newcommand{\exv}[1]{\langle{#1}\rangle}

\numberwithin{equation}{section}

\def\lt#1{\left#1}
\def\rt#1{\right#1}
\def\t#1{\widetilde{#1}}

\def\frc#1#2{\frac{#1}{#2}}

\newcommand{\p}{\partial}

\newcommand{\bra}{\langle}
\newcommand{\ket}{\rangle}

\newcommand{\ii}{ {\rm i} }
\usepackage{authblk}

\newcommand{\varep}{\varepsilon}

\newcommand{\ww}{\boldsymbol{w}}

\newcommand{\bb}{\boldsymbol{ \beta}}
\newcommand{\pbb}{\partial_x\boldsymbol{ \beta}}

\newcommand{\ri}{{\rm i}}
\newcommand{\dd}{{\rm d}}

\title{Diffusive Hydrodynamics of Inhomogenous Hamiltonians  }
\usepackage{authblk}\author[1]{Joseph Durnin}
\author[2]{Andrea De Luca}
\author[2]{Jacopo De Nardis}
\author[1]{Benjamin Doyon}
\affil[1]{ Department of Mathematics, King's College London, Strand WC2R 2LS, London, U.K.}
\affil[2]{ Laboratoire de Physique Théorique et Modélisation, CNRS UMR 8089,CY Cergy Paris Université, 95302 Cergy-Pontoise Cedex, France.}
\date{}                     
\begin{document}

\maketitle

\begin{abstract}
We derive a large-scale hydrodynamic equation, including diffusive and dissipative effects, for systems with generic static position-dependent driving forces coupling to local conserved quantities. We show that this equation predicts entropy increase and thermal states as the only stationary states. The equation applies to any hydrodynamic system with any number of local, PT-symmetric conserved quantities, in arbitrary dimension. {It is fully expressed in terms of elements of an extended Onsager matrix.} {In integrable systems, this matrix admits an expansion in the density of excitations. We evaluate exactly its 2-particle-hole contribution, which dominates at low density, in terms of the scattering phase and dispersion of the quasiparticles, giving a lower bound for the extended Onsager matrix and entropy production.} We conclude with a molecular dynamics simulation, demonstrating thermalisation over diffusive time scales in the Toda interacting particle model with an inhomogeneous energy field.
\end{abstract}

 \tableofcontents

\section{Introduction}
\label{sec:intro}

Interacting systems comprised of many elementary constituents are notoriously hard to describe. The theory of hydrodynamics provides one significant approach to reducing the apparent complexity of the dynamics of such systems, by restricting the descriptors of the dynamics to a smaller set, whose evolution is given in terms of non-linear partial differential equations. In the past few years there has been a resurgence in the use of hydrodynamic techniques. Prominent examples can be found in the study of gravity and relativistic fluids \cite{Rangamani2009,deBoer2015}, strongly coupled field theories \cite{Lucas2015} and electron gases  \cite{Lucas2018,Ku2020,PhysRevLett.118.226601}. Specialising to many-body theory in one spatial dimension, the theory of generalised hydrodynamics (GHD) \cite{PhysRevX.6.041065,PhysRevLett.117.207201}, together with numerous subsequent works \cite{SciPostPhys.2.2.014,PhysRevLett.125.240604,vir1,PhysRevB.101.180302,2005.13546,Gopalakrishnan2019,PhysRevB.102.115121,PhysRevB.96.081118,PhysRevLett.125.070601,PhysRevLett.124.210605,PhysRevLett.124.140603,Bulchandani_2019,SciPostPhys.3.6.039,Doyon_2017}, has represented a major breakthrough in attempts to describe the non-equilibrium dynamics of real-life strongly correlated gases of bosons and low-dimensional magnets \cite{PhysRevLett.122.090601,2009.06651,2006.08577,2009.13535}. 

The conceptual approach underpinning the theory of hydrodynamics is the separation of scales \cite{ldlandau2013}, whereby the microscopic system is coarse-grained into mesoscopic fluid cells, wherein local relaxation is assumed to occur over mesoscopic timescales. Hydrodynamics then describes the evolution, on macroscopic space-time scales, of the parameters characterising the local states that the system has relaxed to. According to the hydrodynamic principle, these parameters are the chemical potentials associated to the local and quasi-local conserved quantities of the dynamics, and the hydrodynamic equations follow by imposing the conservation laws associated to these conserved quantities. The states of most apparent relevance are the maximal entropy states, states defined by the properties of being steady, homogeneous, and ergodic under the dynamics. Taking these as the local states, we obtain the Euler-scale equations: the equations of motion for the convective flows of all conserved quantities. By adding more detailed information on the spatial modulation of the local states, it is possible to include the effect of dissipation and viscosity by adding diffusive terms to the Euler-scale equations. Inclusion of these terms is usually associated with entropy production: the configuration space corresponding to the microscopic scales broadens as large-scale structures are smoothed out. Thus, each fluid cell sees its entropy increase, an effect which is absent in the Euler-scale description. In conventional fluids, the Euler equation with the addition of viscosity effects is known as the Navier-Stokes equation. 

Other terms in the Euler and Navier-Stokes equations are often included to describe the effect of external fields, such as gravitational fields in the original formulation of the Euler and Navier-Stokes equations. These are so-called force or acceleration terms, often resulting from coupling to the density of particles or energy in the fluid. For Galilean (relativistic) invariant  systems it is  straightforward to add a term representing a coupling to the number (energy) to the Navier-Stokes equation. This is because a special simplification occurs: the current of  particles  (energy) is  the  momentum  density,  which is itself a conserved quantity. However, the diffusive hydrodynamics in the presence of generic forces, as in magneto- and thermo- hydrodynamics \cite{Davidson2001} and electron gases in magnetic fields \cite{PhysRevLett.118.226601,PhysRevX.10.011019}, does not contain such a simplification. 

In this paper we derive the general, multi-component diffusive hydrodynamic equations accounting for generic force fields. This generalises the Euler-scale results of \cite{SciPostPhys.2.2.014,Doyon2021} to the diffusive order. We express all terms using appropriate Onsager coefficients, written as time-integrated correlation functions of generalised currents as in the Green-Kubo formula. We use the quantum microscopic dynamics, involving the Kubo-Mori-Bogoliubov inner products coming from perturbation theory and the Kubo-Martin-Schwinger relations. However, by standard arguments, the final results apply equally well to quantum and classical systems. The rate of entropy production which we obtain is shown to be non-negative, and generalises the known Onsager expression for the entropy rate under charge gradients \cite{PhysRev.37.405,PhysRev.38.2265}. { We show that thermal states of the inhomogeneous evolution hamiltonian, accounting for the force fields and with arbitrary chemical potentials for ultra-local quantities (such as the total mass in Galilean systems), are stationary under our hydrodynamic equation.} We mostly focus on systems in one spatial dimension, however the derivation is easily extended to higher dimensions and we give the final fluid equation in any dimension.

Our results have notable implications for the dynamics of integrable one-dimensional systems subjected to external fields. Integrable systems are characterised by the presence of a large number of local and quasi-local conserved quantities, which provide dynamical constraints prohibiting conventional thermalisation. Many experimental and theoretical results have emerged which have confirmed the validity of the following simple principle: isolated purely integrable Hamiltonian systems, both classical and quantum, relax to statistical ensembles which are obtained by maximising entropy under the constraints provided by the full set of conserved quantities present in the system. These are the so-called generalised Gibbs ensembles (GGEs) \cite{Eisert2015,rigol2007relaxation,PhysRevLett.106.227203,Ilievski_2016_str,PhysRevLett.115.157201,Vernier_2017,Bastianello_2017,Essler_2017}. The correct hydrodynamical description of such systems is GHD, where all such conserved quantities are taken into account.

It is generally expected that the addition of terms in the Hamiltonian that break all but a few of the conserved charges should restore canonical thermalisation. However, the old numerical experiment of Fermi-Pasta-Ulam-Tsingou \cite{Dauxois2008} and the more recent cold-atom experiment of the quantum Newton's cradle \cite{Kinoshita2006}, left no doubts that one-dimensional interacting systems of particles whose Hamiltonian dynamics are almost integrable can fail to thermalise on very large time-scales, prompting many theoretical and experimental investigations of the topic \cite{Langen2013,Kitagawa_2011,Gring2012,PhysRevLett.115.180601,Gring1318,Mallayya2019,PhysRevLett.119.010601,2007.01286,PhysRevResearch.2.022034,Langen2016,PhysRevX.8.021030,2103.11997,Biella2016,Biella2019}.

Thus for integrable systems, as potentials coupling to the local and quasi-local conserved charges of the system generically break integrability, we expect that a hydrodynamic description should describe the approach to thermal stationary states. This can reproduce for example the effect of external trapping potentials in cold atomic systems, different interactions with external fields, or spatial inhomogeneities in the system's Hamiltonian. While Euler hydrodynamics has thermal states among its stationary states, the equations do not allow for the entropy production required to reach this state from a generic initial condition \cite{SciPostPhys.2.2.014,Doyon2021}. The inclusion of viscosity terms is therefore fundamental to provide the required entropy production and the approach to thermalisation. This was already shown in the simplest case of a coupling to the density \cite{PhysRevLett.125.240604} in a Galilean invariant system, where the analysis simplifies considerably as mentioned above. Here we consider generic classical and quantum integrable systems and generic forces, { and derive consistent hydrodynamic equations describing the approach to thermalisation, which fully justify the simplified equation used in \cite{PhysRevLett.125.240604} and give an explicit lower bound for entropy production.}

\subsection{Presentation of the problem and main result}

We consider an isolated quantum system with $n$ (which is infinite in integrable systems) extensive conserved quantities in involution,
\begin{equation}
    Q_i = \int dx \ q_i(x)
\end{equation}
with $[Q_i,Q_j]=0$ for all $i,j$. These are conserved in the sense that they are invariant with respect to any Hamiltonian formed of linear combinations of $Q_i$'s. The Hamiltonian we choose below is inhomogeneous, under which they are not necessarily conserved, however as inhomogeneity length scales are large, these conserved quantities play a role within the emergent hydrodynamic description.

The charge densities $q_i(x)$ either act non-trivially on finite regions surrounding $x$ (local) or have norms that decay quickly enough away from $x$ (quasi-local). The charge densities are assumed to be hermitian operators $q_i^\dagger = q_i$. We further assume that there exists a parity and time (PT) inversion symmetry under which densities transform simply: an anti-unitary algebra involution such as 
\begin{equation}
  \mathcal{PT} (q_i(x))   = q_i(-x) , 
\end{equation}
As a consequence, the charges $Q_i$ are PT invariant.  We also introduce the current operators $j_{k,i}(y)$ defined by the flow induced by each conserved quantity, as \cite{SciPostPhys.2.2.014}
\begin{equation}\label{eqmotion1}
\ii [Q_k, q_i(x)] + \partial_x j_{k,i}(x)=0.
\end{equation}
The currents can be chosen hermitian, and by the assumed PT invariance, they can be chosen to transform simply under the PT symmetry, $\mathcal{PT}(j_{k,i}(x)) = j_{k,i}(-x)$. By anti-unitarity $\mathcal{PT}(q_i(x,t)) = q_i(-x,-t)$ and $\mathcal{PT}(j_{k,i}(x,t)) = j_{k,i}(-x,-t)$.
Our results are valid for a generic number $n$ of conserved quantities and we shall later apply it to the integrable case where $n$ is infinite. We shall first restrict ourselves to one spatial dimension and later generalise the result to higher dimensions.

Applying external fields coupling to the densities, the most generic inhomogenous Hamiltonian reads
\begin{equation}\label{eq:Hamiltonian}
H = \sum_{i=1}^n \int dx \  w^i(x) q_i(x),
\end{equation}
where $w^i(x)$ are generic functions of $x$. The time evolved observables $o(x,t)$ are denoted as usual
\begin{equation}
    o(x,t) = e^{\ii t H}o(x)e^{-\ii tH}.
\end{equation}

We assume that the $w^i(x)$ vary slowly in space such that we can expand them around any point $x_0$ as 
\begin{equation}
      w^i(x) =  w^i(x_0) + (x-x_0)\partial_{x_0} w^i(x_0) + \ldots,
\end{equation}
with small higher order corrections. The accuracy of the resulting hydrodynamic equations is determined by how small such derivatives are with respect to the miscroscopic scales of the model; we leave a precise analysis for future works, but the example of the Toda gas in section \ref{sec:toda} will give some intuition. Effectively, the system is subject to external inhomogenous forces 
\begin{equation}
     \mathfrak{f}^i(x) = - \frac{\partial w^i(x) }{\partial x}
\end{equation}
which locally break the conserved quantities. In the following we will use the repeated indices convention to denote sums over indices. 

Our main result is a hydrodynamic equation for the space-time evolution of the local expectation values of the charge densities 
\begin{equation}
    {\tt q}_i(x,t) = \langle q_i(x,t)\rangle_{\rm ini},
\end{equation}
with respect to some initial state $\langle \cdots\rangle_{\rm ini}$, up to second order in spatial derivatives. This includes diffusive terms which are responsible for entropy increasing and thermalisation. We define $\ell = {\rm min}_i(|\p_x {\tt q}_i(x,t)|^{-1})$, the spatial scale of variation of the local densities which, in the hydrodynamic approximation, determines the scale of variation of other local observables, and $\ell_{\mathfrak f} = {\rm min}_i(|\mathfrak f^i|^{-1})$. The resulting hydrodynamic equations, around the point $x,t$, are correct up to and including terms of order $1/\ell^2,\, 1/\ell_{\mathfrak f}^2, 1/(\ell\ell_{\mathfrak f})$. With increasing time, $\ell$ is expected to increase at almost all points, up to the scale $\ell_{\mathfrak f}$ determined by the external fields, with the possible exception of isolated points where shocks or other singular structures may develop. Therefore at large enough times, we assume $\ell\approx \ell_{\mathfrak f}$.

It is convenient to characterise the state at $x,t$ by a local maximal entropy state (a local Gibbs of generalised Gibbs ensemble, which we will simply refer to as a local GGE) which reproduces the averages ${\tt q}_i(x,t)$. For this purpose, we define the thermodynamic potentials $\beta^i(x,t)$ by inverting the defining relation 
\begin{equation}\label{eq:hydrostate}
   {\tt q}_i(x,t)  = \frac{{\rm tr}[ q_i e^{-   Q_l \beta^l(x,t)}]}{{\rm tr}[   e^{-   Q_l \beta^l(x,t)}]},
\end{equation}
at each position $x,t$. We denote by $\langle\cdots\rangle$ the GGE with thermodynamic potentials $\beta^i(x,t)$, where here and throughout we keep the $x,t$ dependence implicit. This forms a family, parametrised by $x,t$, of homogeneous and stationary states on the infinite line.

Let us consider a single fluid cell at position $x,t$, and introduce several quantities associated to it.
Given any $n$-dimensional vectors $\boldsymbol{a},\;\boldsymbol{b}$ (which may depend on $x,t$), we shall use the following compact notation for the contraction of indices with the external fields or forces
\begin{equation}
j_{ \boldsymbol{ a},k}(y)= a^i   j_{i,k}(y), \quad j_{i,\boldsymbol{b}}(y)=  j_{i,k}(y)b^k,\quad j_{ \boldsymbol{ a},\boldsymbol{ b}}(y)= a^ij_{i,k}(y)b^k.
\end{equation}
We denote the time evolution of a generic operator $o(y)$ in complex time as
\begin{equation}\label{eq:evolwithw}
o(y, s \boldsymbol{w} - \ii \tau \boldsymbol{\beta} ) = e^{\ii s w^i Q_i + \tau \beta^l Q_l} \ o(y) \ e^{-\ii s w^i Q_i - \tau \beta^l Q_l}
\end{equation}
where $w^i=w^i(x)$ and $\beta^i=\beta^i(x,t)$. Within the formulae below, this is to be interpreted as microscopic time evolution, occurring within the mesoscopic cell at $x,t$. In particular, the real part of the time evolution is with respect to the Hamiltonian $H_{x} =   w^i(x) Q_i$, which is \eqref{eq:Hamiltonian} taken with the fields constant, at their local value at $x$. We also define the generalised Kubo-Mori-Bogoliubov (KMB) inner product for two local operators in terms of the connected correlation function
\begin{equation}\label{genKMB}
    (o_1(y,s\boldsymbol{w}),o_2(0)) =
    \int_0^1 
    d\lambda\,
    \langle{o_1(y,s \boldsymbol w -\ii\lambda\boldsymbol{\beta} )o_2(0)}\rangle^{\rm c}.
\end{equation}
Throughout we shall require that the inner product is sufficiently clustering, specifically that $(o_1(y,s\ww),o_2(0,0))$ decays faster than $1/|y|$ at large $|y|$. We can then introduce the following extended Onsager coefficients for a generic local operator $o(x)$, 
\begin{align}\label{eq:def_Onsager}
& \mathfrak{L}[\boldsymbol{ a}, \boldsymbol{ b} ;o] =        \int_{-\infty}^{\infty} ds \,    (     J_{ \boldsymbol{ a},\boldsymbol{ b}}(s \ww   ),     o )^C   ,
\end{align}
with $J_{i,k}(s\ww) = \int_{-\infty}^\infty dy \ j_{i,k}(y,s\ww)$, the spatially integrated current,  and $o=o(0)$. Here we have introduced the connected inner product, denoted $^C$, which is based on clustering at large times:
\begin{align}\label{eq:connected}
&  ( O_1(s\ww), o_2 )^C =      (  O_1(s\ww) , o_2 )  -  \lim_{s \to \infty}   (  O_1(s\ww) , o_2 ).
\end{align}
By the method of hydrodynamic projections, we can write the temporally disconnected component as
\beq\label{eq:projection}
\lim_{s \to \infty}   (  O_1(s\ww) , o_2 )
= (  \mathbb P O_1 , o_2 )= \sum_{ij} (O_1,q_i)\mathsf C^{ij}(Q_j,o_2),
\eeq
where $\mathsf C^{ij}$ is the inverse of the susceptibility matrix ${\mathsf C}_{ij} = \langle Q_i q_j \rangle^c$, and $\mathbb P$ is the projector onto the space of conserved quantities $Q_i$. In terms of explicit double-indices $(i,j)$ and $(k,l)$, the extended Onsager matrix $\mathfrak L$ has matrix elements $\mathfrak L_{i,j;k,l}\equiv \mathfrak L[i,j;j_{k,l}]$ defined by
\beq\label{OnsagerMatrix}
    a^ib^jc^kd^l \mathfrak L_{i,j;k,l}
    = \mathfrak L[\boldsymbol a,\boldsymbol b;j_{\boldsymbol c,\boldsymbol d}].
\eeq
\begin{figure}
    \centering
    \includegraphics[width=0.7\textwidth]{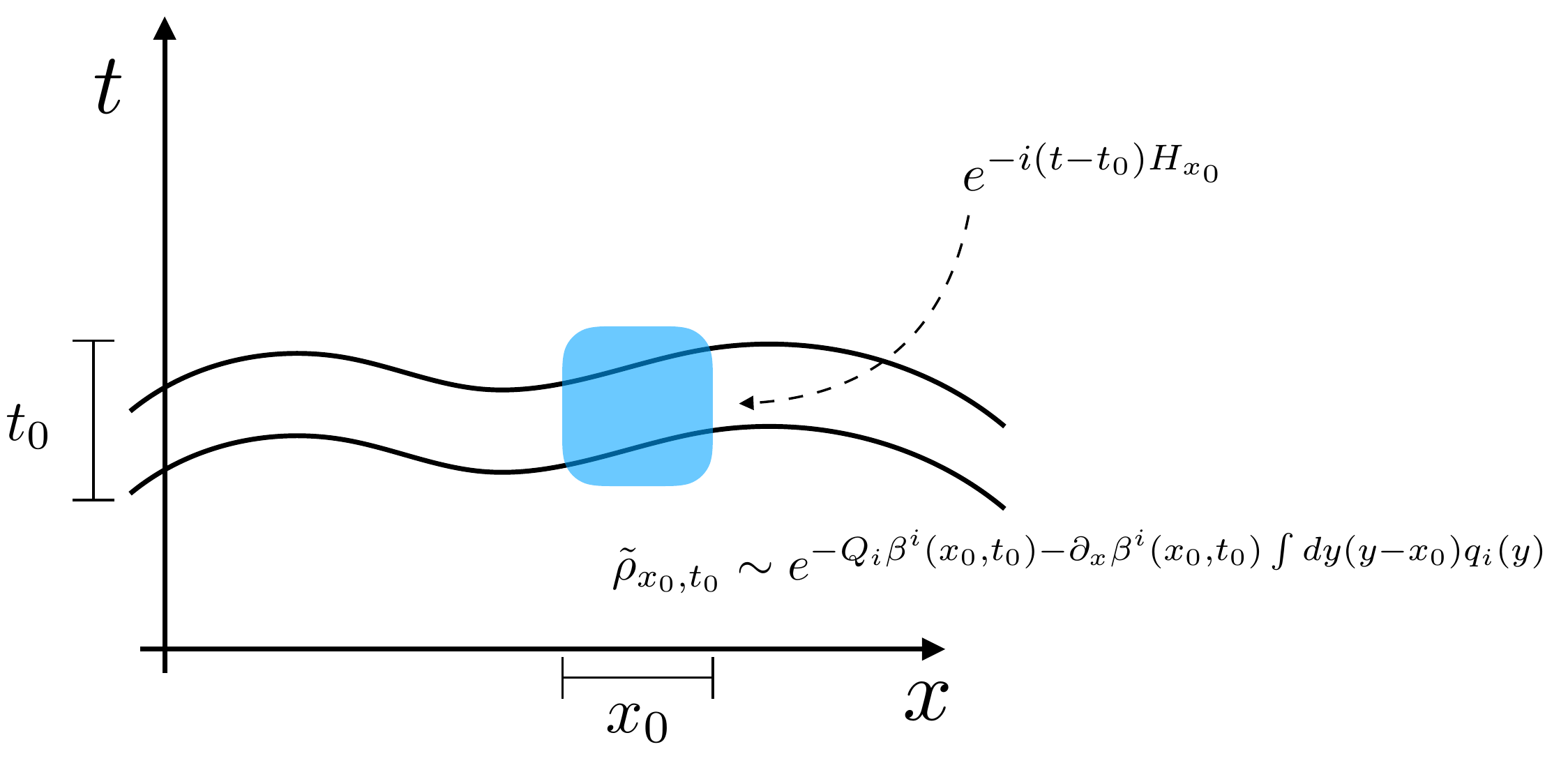}
    \caption{Graphical representation of the derivation of main result eq. \eqref{main_result}. At each mesoscopic time $t_0$, the local hydrodynamic state is expanded in the slow variation of the chemical potentials, eq. \eqref{eq:localstatx0}. The state is then evolved with the local Hamiltonian \eqref{eq:localHx0}, consisting of a global conservative part, and the force term, which breaks most of the conservation laws. This time evolution describes local relaxation, and thus occurs purely within the fluid cell. The outputs of this process are expressions for expectation values of currents within the fluid-cell, expressed in terms of fluid-cell expectation values of densities. These are then inserted into the Heisenberg equations of motion to obtain the hydrodynamic equation.}
    \label{fig:figu}
\end{figure}
We now proceed to write the hydrodynamic equation governing the evolution of the local expectation values ${\tt q}_i(x,t)$. 
The derivation of our main result is provided in detail in later sections. The main steps required are graphically shown in Fig.~\ref{fig:figu}, and can be summarised as follows:
\begin{enumerate}
 \item The evolution of the densities $q_i$ and currents $j_i$ at position $x_0$, governed by the local Hamiltonian 
 \begin{equation}\label{eq:localHx0}
     H_{x_0}= w^i(x_0) Q_i -  \mathfrak{f}^i(x_0) \int dx\, (x-x_0)   q_i(x),
 \end{equation}
  are obtained using perturbation theory. The form of the Hamiltonian is obtained by expanding \eqref{eq:Hamiltonian} to first order in $\mathfrak{f}^i$ about the point $x_0$.
    \item The resulting expectation values of the charges $q_i$ and currents $j_i$, evolved using perturbation theory, are expanded to first order in $\partial_{x_0}\beta^i(x_0,t_0)$ within the hydrodynamic cell at $x_0,t_0$. Each cell has the local state
    \begin{equation}\label{eq:localstatx0}
        \tilde{\rho}_{x_0,t_0} \propto e^{-  Q_i \beta^i(x_0,t_0) - \partial_{x_0}\beta^i(x_0,t_0) \int d y (y-x_0) q_i(y) }.
    \end{equation}
    \item By inverting the resulting expressions for the perturbatively time-evolved, fluid-cell expanded expectation values, which are written in terms of $\beta^i(x_0,t_0)$ and $\partial_{x_0}\beta^i(x_0,t_0)$, we express the expectation values of currents in terms of expectation values of densities (and their spatial derivatives). Here, the expectation values of densities are the ``finite-time" hydrodynamic variables.
    \item We take the large time $t$, see Fig. \ref{fig:figu}, limit of these expressions, which incorporates the effect of local relaxation. This gives the currents in terms of the true hydrodynamic variables in the fluid cell at $x_0,t_0$: the ${\tt q}_i(x_0,t_0)$ and their spatial derivatives $\partial_{x_0}{\tt q}_i(x_0,t_0)$. The mapping between these and the potentials is provided by \eqref{eq:hydrostate}. 
    \item We insert the resulting expressions for the currents into the Heisenberg equation of motion $\p_t q_i = \ri [H,q_i]$. Using the density-density commutation relation \eqref{commutatorqq}, shown in
    \cite{SciPostPhys.2.2.014,Doyon2021}, this can be written in terms of the microscopic currents:
    \begin{equation}\label{continuity2}
        \partial_t q_i( x,t) =  - \partial_x j_{\ww,i} (x,t  ) +  j_{i,\boldsymbol{\mathfrak{f}}}(x,t).
    \end{equation}
    When written for the hydrodynamic variables, $q_i(x,t)\to {\tt q}_i(x,t)$, this becomes an equation for the evolution of local states.
\end{enumerate}
We stress that even if the expansion of the state and of the Hamiltonian are taken only to first order in $\partial_x$, the hydrodynamic equation resulting from this procedure is correct up to second order in spatial derivatives, because all terms on the RHS of \eqref{continuity2} are already first order in derivatives. Suppressing the $x,t$ dependence of each term, the full hydrodynamic equation reads:
\begin{align}\label{main_result}
&\partial_t {\tt q}_i + \partial_x {\tt j}_{\boldsymbol{w},i}  +  \frac{1}{2}\partial_x\left(\mathfrak{L}[\boldsymbol{w}, \partial_x\boldsymbol{ \beta} ; j_{\boldsymbol{w},i}]  +  \mathfrak{L}[\boldsymbol{\beta}, \boldsymbol{\mathfrak{f}} ; j_{\boldsymbol{w},i}]\right) \nonumber \\=&\  {\tt j}_{i,\boldsymbol{\mathfrak{f}} } + \frac{1}{2}\left( \mathfrak{L}[\boldsymbol{w}, \partial_x \boldsymbol{  \beta} ; j_{ i,\boldsymbol{\mathfrak{f}}  }] +  \mathfrak{L}[\boldsymbol{\beta}, \boldsymbol{\mathfrak{f}} ; j_{i,\boldsymbol{\mathfrak{f}} }] \right)
\end{align}
where we denoted the Euler expectation values of currents on local macroscopic states as 
\begin{equation}\label{eq:jGGE}
     {\tt j}_{i,k}(x,t) = \frac{{\rm tr}[ j_{i,k} e^{-   Q_l \beta^l(x,t)}]}{{\rm tr}[   e^{-   Q_l \beta^l(x,t)}]}.
\end{equation}
Notice that the spatial derivatives of the thermodynamic potentials are related to the spatial derivatives of the expectation values of the charges. In terms of the susceptibility matrix  ${\mathsf C}_{ij}$, this reads
\begin{equation}
{\mathsf C}_{ij} \partial_x {\beta}^j = - \partial_x  {\tt q}_i .
\end{equation}
{ Eq.~\eqref{main_result} gives an unbroken continuity equation for the inhomogeneous energy density $w^i(x){\tt q}_i(x)$, which follows by linearity of the current operator $j_{i,\boldsymbol a}$ as a function of ${\boldsymbol a}$.  \red{Using the relation $w^i(x) \partial_x\mathfrak{L}[\boldsymbol{w}, \partial_x\boldsymbol{ \beta} ; j_{\boldsymbol{w},i}] =  \partial_x\mathfrak{L}[\boldsymbol{w}, \partial_x\boldsymbol{ \beta} ; j_{\boldsymbol{w},\boldsymbol{w}}] + \mathfrak{L}[\boldsymbol{w}, \partial_x\boldsymbol{ \beta} ; j_{\boldsymbol{w},\boldsymbol{\mathfrak{f}} }]$, and the analogous relations for other terms, we see}        
\beq
\partial_t (w^i{\tt q}_i) +   \partial_x[ {\tt j}_{\boldsymbol{w},\boldsymbol w}  + \frac{1}{2} \left(\mathfrak{L}[\boldsymbol{w}, \partial_x\boldsymbol{ \beta} ; j_{\boldsymbol{w},\boldsymbol w}]  +  \mathfrak{L}[\boldsymbol{\beta}, \boldsymbol{\mathfrak{f}} ; j_{\boldsymbol{w},\boldsymbol w}]\right)]   =0.
\eeq
Eq.~\eqref{main_result} also returns conservation laws for any {\em ultra-local} density. This is defined as a density $u^i q_i(x)$, for some constants $u^i$, whose charge does not generate any current, $u^i j_{i,k}(x)=0$. An equivalent definition of an ultra-local density is $u^i[Q_i, q_k(x)]=0$, for all $k$. In this case we have again the continuity equation
\beq
\partial_t (u^i {\tt q}_i) +    \partial_x[ {\tt j}_{\boldsymbol{w},\boldsymbol u} + \frac{1}{2}  \left(\mathfrak{L}[\boldsymbol{w}, \partial_x\boldsymbol{ \beta} ; j_{\boldsymbol{w},\boldsymbol u}]  +  \mathfrak{L}[\boldsymbol{\beta}, \boldsymbol{\mathfrak{f}} ; j_{\boldsymbol{w},\boldsymbol u}]\right) ] =0.
\eeq
Examples of ultra-local densities are the particle density in Galilean models, and the magnetic field in spin and Fermi-Hubbard chains. 
}

The usual case, with a dynamics induced by a homogeneous Hamiltonian, is recovered by setting $\mathfrak{f}=0$, hence $ {\tt j}_{i,\boldsymbol{\mathfrak{f}} }=0$ and therefore
\begin{equation}\label{eq:simplify}
    \mathfrak{L}[\boldsymbol{a}, \boldsymbol{ b} ; j_{i,\boldsymbol{\mathfrak{f}}}]=\mathfrak{L}[ \boldsymbol{ a}, \boldsymbol{\mathfrak{f}} ; o] =0.
\end{equation}
%
The simplification \eqref{eq:simplify} also holds whenever the force only couples to a conserved density, such as $q_0(x)$ (that is, $\mathfrak f^k = \delta^k_0$), whose currents with respect to all flows, i.e. $j_{k,0}(x)$ for all $k$, are themselves conserved densities. We will refer to such a $q_0(x)$ as \textit{self-conserved}. In this case, as conserved quantities are projected out in \eqref{eq:connected}, the resulting Onsager coefficients vanish. This phenomenon is observed, for instance, for the local particle density in integrable or non-integrable Galilean systems, where we have $j_{k,0} = q_{k-1}$ $\forall$ $k$ by Galilean invariance, in a suitable basis for the charges. Similar statements hold for relativistic models, and for the XXZ chain, where in both cases the energy density is self-conserved. Equalities of this type, for averages in a GGE, arise from the thermodynamic Bethe ansatz and GHD expressions, as first noted in \cite{SciPostPhys.2.2.014}; at the level of operators, they are related to the observation that the boost operator preserves the integrable hierarchy \cite{10.21468/SciPostPhys.8.2.016}.

Another simplification occurs if the external field $\mathfrak f^kq_k(x)$ is ultra-local. In this case the Onsager coefficients $\mathfrak{L}[\boldsymbol{w}, \partial_x\boldsymbol{ \beta} ; o]$ become those taken with respect to the homogenenous background Hamiltonian with $\boldsymbol w$ constant. The case of coupling to the particle density in Galilean systems, considered in \cite{PhysRevLett.125.240604}, admits both the aforementioned simplifications. Note that the stronger requirement that all $j_{k,0}$ be themselves conserved densities, for the simplification \eqref{eq:simplify}, was missed in \cite{PhysRevLett.125.240604}.

Finally, we note that, as the above overview of the derivation suggests, two main timescales are involved, which are assumed to be well separated: the local thermalisation time to relax in the local hydrodynamic cell (local relaxation, assumed much faster) and the (longer) hydrodynamic large-scale dynamics; see \eqref{times} and \eqref{timesforce}. In the latter the presence of force terms can induce a global thermalisation to a thermal Gibbs ensemble where only energy and few other quantities describe the state, as explained below. This dynamics is on diffusive timescales and regulated by the strength of diffusion terms and force terms. In our construction, these are the only timescales which can enter the problem.

\subsection{Entropy increase and stationarity}
The equation \eqref{main_result} guarantees positive thermodynamic entropy increase and thermal states as the only stationary states of the evolution. The definition of the entropy density leads to the time-evolution
\begin{equation}
\partial_t s(x,t)  = \beta^i \partial_t   {\tt q}_i(x,t)  .
\end{equation} 
Using that the Euler part of the hydrodynamic equation does not lead to entropy increase \cite{Doyon2021},
\begin{equation}
    \int dx \,\beta^i ( \partial_x {\tt j}_{\boldsymbol{w},i} - {\tt j}_{i,\boldsymbol{\mathfrak{f}} })=0,
\end{equation}
and spatially integrating \eqref{main_result} by parts, the increase in the total entropy $S(t)= \int dx \, s(x,t)$, is given by the following combination of extended Onsager coefficients: 
\begin{align}\label{eq:entropyIncrease}
&\partial_t S = \frac{1}{2} \int_{-\infty}^{\infty} dx\,\Big( \mathfrak{L}[\boldsymbol{w}, \partial_x\boldsymbol{  \beta} ; j_{\boldsymbol{w},\partial_x\boldsymbol{  \beta}}]  + \mathfrak{L}[ \boldsymbol{  \beta}, \boldsymbol{\mathfrak{f}} ; j_{\boldsymbol{w},\partial_x\boldsymbol{  \beta}}] + \mathfrak{L}[\boldsymbol{w}, \partial_x\boldsymbol{ \beta} ; j_{\boldsymbol{  \beta},\boldsymbol{\mathfrak{f}}}] + \mathfrak{L}[\boldsymbol{  \beta}, \boldsymbol{\mathfrak{f}} ; j_{\boldsymbol{  \beta},\boldsymbol{\mathfrak{f}}}]\Big).
\end{align}
Using the definition of the extended Onsager coefficients \eqref{eq:def_Onsager}, the entropy increase takes the quadratic form
\begin{align}\label{eq:Squadratic}
&\partial_t S = \frac{1}{2} \int_{-\infty}^{\infty} dx \int_{-\infty}^{\infty} ds\, (J_{\ww,\pbb}(s \ww  ) + J_{\bb,\boldsymbol{\mathfrak{f}}}(s \ww ),j_{\ww,\pbb }+ j_{\bb,\boldsymbol{\mathfrak{f}}})^C .
\end{align}
The right-hand-side is always non-negative
\beq\label{eq:positiventropy}
    \p_t S \geq 0,
\eeq
as the KMB inner product $(\cdot,\cdot)$ is positive semi-definite\footnote{In fact, it is in general a pre-inner product, as it is not necessarily positive-definite.}. Therefore $(O,o)$ is non-negative, as by translation invariance and clustering it can be written as the limit
\beq\label{eq:positiveOo}
    (O,o) = \lim_{L\to\infty}\frc1{2L}\left(\int_{-L}^L dy\, o(y)\;,\;  \int_{-L}^L dz\,o(z)\right)\geq 0.
\eeq
Then, $(O,o)^C = ((1-\mathbb P)O,(1-\mathbb P)o)\geq 0$, and finally, by stationarity and clustering at large times,
\begin{equation}
\int_{-\infty}^\infty ds\,( O(s \ww ) , o )^C = \lim_{T\to\infty}
\frc1{2T} \left( \int_{-T}^T ds\, O(s \ww),\int_{-T}^T du\, o(0,u \ww)\right)^C\geq 0,
\end{equation}
showing \eqref{eq:positiventropy}.
It should be stressed that equation \eqref{eq:entropyIncrease} generalises the entropy production rate induced by thermodynamic forces found by Onsager \cite{PhysRev.37.405,PhysRev.38.2265} (see also more recent works \cite{PhysRevResearch.2.022009,PhysRevLett.115.090601,PhysRevE.101.012132}) to a system with $n$ conserved quantities and external inhomogeneous force fields. Analogously to the work of Kubo \cite{Kubo1957} we here prove positivity of entropy increase by the definition of the extended Onsager coefficients. { In particular, by the above discussion the extended Onsager matrix define in \eqref{OnsagerMatrix} is positive semi-definite,
\beq
    \mathfrak L \geq 0
\eeq
and the entropy production formula is
\beq
    \p_t S = \frc12 m^{i,j} \\ \mathfrak L_{i,j;k,l}  m^{k,l}   ,\quad
    m^{i,j} = w^i \p_x \beta^j + \beta^i \mathfrak f^j.
\eeq
}

As entropy can only increase, stationarity should be reached when the entropy of local states is maximal. The condition of stationary entropy is obtained from \eqref{eq:Squadratic} by considering the non-negativity result \eqref{eq:positiveOo}. It implies that we must have
\beq
\int_{-\infty}^\infty ds\,(J_{\ww,\pbb}(s \ww  ) + J_{\bb,\boldsymbol{\mathfrak{f}}}(s \ww ),j_{\ww,\pbb }+ j_{\bb,\boldsymbol{\mathfrak{f}}})^C = 0
\eeq
at every point $x$. Suppose that there is some conserved density $q_0$ which is ultra-local and self-conserved, and $q_1$, which is ultra-local but not necessarily self-conserved. Then a family of stationary entropy solutions is given by
\beq\label{eq:conditionSmax}
    \beta^i(x) = \beta ( w^i(x) - \mu_0(x)  \delta_0^i -  \mu_1  \delta_1^i),
\eeq
where $\beta$ and $\mu_1$ are constants, $\mu_0(x)$ is an arbitrary function of $x$, and $\boldsymbol e_{0,1}$ are the associated basis vectors. 

However, a stationary entropy does not necessarily guarantee that a stationary solution has been reached, as it only accounts for stationarity under the diffusive terms in the hydrodynamic equation. The condition \eqref{eq:conditionSmax} makes all Onsager terms in \eqref{main_result} cancel, however there generally remains a nontrivial Euler-scale evolution, which preserves entropy. The fully stationary solutions are those where $\partial_x {\tt j}_{\boldsymbol{w},i} = {\tt j}_{i,\boldsymbol{\mathfrak{f}} }$, which imposes the constraint that $\mu_0(x)$ is a constant in $x$ \eqref{eq:conditionSmax} \cite{SciPostPhys.2.2.014,CauxCradle,Doyon2021}. If it is not, the resulting Euler-scale evolution exits the space of states \eqref{eq:conditionSmax}, and diffusion then further increases entropy. At large times, the solution reached is then for $\mu_0(x)=\mu_0$, a constant. This mechanism is described in \cite{PhysRevLett.125.240604}. This shows that the expected thermal (Gibbs) states
\beq\label{LDA_state}
\beta^i(x) = \beta (w^i(x) - \mu_0\delta_{0}^i - \mu_1\delta_1^i),
\eeq
where the thermodynamic and external forces cancel out, $\partial_x \beta^i(x) + \beta \mathfrak{f}^i(x)=0$, are stationary states with vanishing entropy production. We note that both the temperature, and the chemical potential associated to any ultra-local conserved charge, are arbitrary parameters of the stationary state. Ultra-local charges are still conserved with an inhomogeneous Hamiltonian, and temperatures and chemical potentials are determined by the initial values of the total energy and of the total charges.


\subsection{Example: few-component hydrodynamics coupled to the energy density}
To give an explicit application of the above, we now consider a Galilean system with conservation of particle number $N=Q_0$, momentum $P=Q_1$ and energy $E=Q_2$. We choose a coupling to the energy field, namely a space-dependent Hamiltonian function,
\begin{equation}
    H = \int d x \,  w(x) h(x).
\end{equation}
This represents the simplest example of a system which requires the full formalism developed in this paper. Specifically we take
\begin{equation}
    w^i(x)=w(x)\delta_2^i,\;\;\mathfrak{f}^i(x)=\mathfrak{f}(x)\delta_2^i.
\end{equation}
The hydrodynamic averages can be written in the following form
\begin{equation}
   {\tt q}_i(x,t)  = \frac{{\rm tr}[ q_i e^{-   T^{-1}(H-\nu P-\mu N)}]}{{\rm tr}[   e^{-   T^{-1}(H-\nu P-\mu N)}]},
\end{equation}
where $T$ is the temperature, $\mu$ the chemical potential, and $\nu$ the boost parameter. We denote the hydrodynamic averages of charge densities ${\tt n}$, ${\tt p}$ and ${\tt e}$, and their respective currents by ${\tt j}_{\tt n}$, ${\tt j}_{\tt p}$, ${\tt j}_{\tt e}$ (and more generally the indices ${\tt n}$, ${\tt p}$, and ${\tt e}$ representing 0, 1, 2). As a consequence of the fact that $j_{{\tt n},i}=0$ and $j_{{\tt p},i}=q_i$, from all extended Onsager matrix elements only the standard Onsager matrix elements remain,
\begin{equation}
    \mathcal{L}_{i;k}=\int_{-\infty}^{\infty} ds \,    (     J_{i}(s \ww   ),    j_k  )^C .
\end{equation}
Further, certain Onsager matrix elements vanish, as a consequence of the fact that ${\tt j}_{\tt n}= {\tt p}$ is a conserved density (the momentum density). We can thus write the full set of hydrodynamic equations explicitly as:
\begin{equation}
    \p_t{\tt n}+\p_x(w{\tt p})=0,
\end{equation}
\begin{align}
    \p_t{\tt p}+\p_x(w{\tt j}_{\tt p})-\frac{\partial_x}{2}\left(\frac{w^2\mathcal{L}_{{\tt p};{\tt p}}}{T}\left(\p_x\nu-\frac{\nu\p_xT}{T}\right)+w^2\mathcal{L}_{{\tt e};{\tt p}}\frac{\p_xT}{T^2} - \frac{w\mathfrak{f}\mathcal{L}_{{\tt e};{\tt p}}}{T}\right)=\mathfrak{f}{\tt e}
\end{align}
\begin{align}
&\partial_t {\tt e} + \partial_x (w{\tt j}_{\tt e})  -  \frac{\partial_x}{2}\left(\frac{w^2\mathcal{L}_{{\tt p};{\tt e}}}{T}\left(\p_x\nu-\frac{\nu\p_xT}{T}\right)+w^2\mathcal{L}_{{\tt e};{\tt e}}\frac{\p_xT}{T^2}  - \frc{w\mathfrak{f}\mathcal{L}_{{\tt e};{\tt e}}}T 
\right) \nonumber \\=&\  \mathfrak{f}{\tt j}_{\tt e} - \frac{1}{2} \left(
\frc{w\mathfrak f \mathcal{L}_{{\tt p};{\tt e}}}T\left(\p_x\nu-\frac{\nu\p_xT}{T}\right)
+ w\mathfrak f \mathcal{L}_{{\tt e};{\tt e}}\frac{\p_xT}{T^2}
- \frac{\mathfrak{f}^2\mathcal{L}_{{\tt e};{\tt e}}}{T}
\right) .
\end{align}
By fundamental thermodynamic identities -- in fact as a consequence of the KMS relation \cite{Doyon2021} -- the currents can be expressed in terms of the potentials, the specific free energy $f$, and some constant $G$ interpreted as the strength of the thermal entropy flux; that is ${\tt j}_{\tt n}= {\tt p} = \nu {\tt n}$, ${\tt j}_{\tt p} = \nu {\tt p} - T f$, and ${\tt j}_{\tt e} = \nu {\tt e} - T\nu f- T^2 G$. Therefore, the hydrodynamic equations up to diffusive scale are fully determined by $f$, $G$ and the three Onsager matrix elements $\mathcal{L}_{{\tt p};{\tt p}}$,  $\mathcal{L}_{{\tt p};{\tt e}} = \mathcal{L}_{{\tt e};{\tt p}}$ and $\mathcal{L}_{{\tt e};{\tt e}}$. If parity symmetry is present then $G=0$.

\subsection{Formulation in $D$ spatial dimensions}

It is a simple matter to extend our calculations and results to higher dimensions. The only technical requirement is that the KMB product satisfy a slighter stronger clustering condition: in $D$-spatial dimensions, $(o_1(\vec{x},t),o_2(0,0))$ decays faster than $1/|\vec{x}|^D$ at large $|\vec{x}|$. Further, the PT transformation is an anti-unitary algebra involution that inverts the direction of all spatial components. We obtain
\begin{align}\label{main_result_highD}
\partial_t {\tt q}_i + \nabla\cdot \vec{{\tt j}}_{\boldsymbol{w},i}  + & \frac{1}{2} \nabla \cdot\left(\mathfrak{L}\left[\boldsymbol{w},  \nabla \boldsymbol{ \beta} ; \vec{j}_{\boldsymbol{w},i}\right]  +  \mathfrak{L}\left[\boldsymbol{\beta}, \vec{\boldsymbol{\mathfrak{f}}} ; \vec{j}_{\boldsymbol{w},i}\right]\right) \nonumber \\& = {\tt j}_{i,\vec{\boldsymbol{\mathfrak{f}} }} + \frac{1}{2}\left( \mathfrak{L}\left[\boldsymbol{w}, \nabla\boldsymbol{  \beta} ; j_{ i,\vec{\boldsymbol{\mathfrak{f}}  }}\right] +  \mathfrak{L}\left[\boldsymbol{\beta}, \vec{\boldsymbol{\mathfrak{f}}} ; j_{i,\vec{\boldsymbol{\mathfrak{f}}} }\right] \right),
\end{align}
where $j_{i,\vec{\boldsymbol{a}} }=\vec{j}_{i,k}\cdot \vec{a}^{\,k}$, and the extended Onsager coefficients in higher dimensions read
\begin{align}\label{eq:def_Onsager_highD}
& \mathfrak{L}[\boldsymbol{ a}, \vec{\boldsymbol{ b }};o] =     \lim_{t \to \infty}     \int_{-t}^{t} ds     \left(     J_{ \boldsymbol{ a},\vec{\boldsymbol b}}(y,s \ww ),     o(0,0) \right)^C   .
\end{align}
Note that $\mathfrak{L}[\boldsymbol{ a}, \vec{\boldsymbol{ b }};o]$ inherit the vectorial type of $o$. For the entropy increase we have similarly:
\begin{align}
&\partial_t S = \frac{1}{2} \int_{-\infty}^{+\infty} ds \left(J_{\ww,\nabla\boldsymbol{  \beta}}(s \ww  ) + J_{\bb,\vec{\boldsymbol{\mathfrak{f}}}}(s \ww ),j_{\ww,\nabla\boldsymbol{ \beta} }(0,0)+ j_{\bb,\vec{\boldsymbol{\mathfrak{f}}}}(0,0)\right)^C \ge0.
\end{align}

\section{Diffusive hydrodynamics with inhomogeneous fields }

The purpose of this section is to derive the main result \eqref{main_result}. The derivation is based on evaluating, from microscopic calculations, an expression for the currents at large times, from an initial condition where thermodynamic potentials are linear in space (constant thermodynamic forces), and under a dynamics where the fields vary linearly in space (constant external forces). The thermodynamic and external forces are taken to be small, and the calculation is perturbative in these forces, with the application of standard perturbation theory. \red{At microscopic scales, there is local relaxation to a state which is near to a maximal entropy state, and which spatially extends to the larger mesoscopic scales.} The expression for the expectation values of currents on these mesoscopic scales, expressed in terms of the charge densities and their derivatives at the same mesoscopic scales, and combined with the conservation laws, is the basis for the hydrodynamic equation governing the macroscopic evolution. The Euler contribution to these current expectation values gives the ``leading order'' of the mesoscopic scales, and diffusive contributions are obtained at sub-leading order.
Similar ideas are used in order to obtain the slow dynamics under integrability breaking in homogeneous situations \cite{2103.11997,Durnin2020,PhysRevB.101.180302,2005.13546}.

\subsection{Diffusive hydrodynamics: constant fields }

In order to illustrate the method, we start with a simpler problem where the external fields are constant in space, where we will obtain the hydrodynamic equation including diffusive effects to the second order in spatial derivatives. Under unitary Heisenberg time evolution and with constant fields $w^i(x)=w^i$ in the Hamiltonian,  eq. \eqref{eq:Hamiltonian}, the conserved quantities $Q_i$ of the system are all exactly conserved. Their densities therefore satisfy a set of continuity equations (denoting with $o(x,t) = e^{i H t }  o(x)  e^{-i H t } $, with $H=w^i Q_i$)
\begin{equation}\label{eq:continuity0}
\partial_t q_i(x,t) +  \partial_x j_{\ww ,i}(x,t) = 0.
\end{equation}
The basic ingredient of the hydrodynamic approach is the hydrodynamic state, which is 
\begin{equation}\label{eq:density_mat}
    \tilde{\rho}_{t} \propto e^{- \int dx\, \beta^l(x,t) q_l(x)  }.
\end{equation}
The state at a given time is thus equivalent to its potentials $\beta^l(x,t)$, viewed as functions of $x$, and hydrodynamics prescribes how to calculate the time-evolution of these potentials. Note however that the potentials appearing in \eqref{eq:density_mat} are not exactly those appearing in \eqref{eq:hydrostate} and \eqref{eq:jGGE}, with the two differing by first-order spatial derivative terms, see the discussion at the end of this subsection.

Crucial to these results is the hydrodynamic approximation, which accounts for two separate effects:

\medskip
\noindent I. Separation into local fluid cells: at every mesoscopic time $t_0$ and position $x_0$, the fluid can be represented by the following state:
\begin{equation}\label{eq:hydro1}
\tilde{\rho}_{x_0,t_0} \propto  e^{ -\beta^l(x_0,t_0) Q_l - \partial_{x_0} \beta^l(x_0,t_0)\, \int dy \: (y-x_0) \: {q}_l(y)  },
\end{equation}
where the functions $\beta(x,t)$ are exactly those appearing in \eqref{eq:density_mat}.
This expression is justified if the state is sufficiently slowly varying in space, and correlations decay sufficiently rapidly in space, as then an expectation value evaluated at some position $x_0$ in the state \eqref{eq:density_mat} will be approximately equal to that in the state \eqref{eq:hydro1}.

\medskip
\noindent II. Local relaxation: hydrodynamic averages ${\tt o}(x_0,t_0)$ in the fluid cell located at $x_0,t_0$ are not defined by equating them to $\bra o\ket_{\t{\rho}_{x_0,t_0}}\equiv\mathrm{Tr}(o\t\rho_{x_0,t_0})$, but instead, they are defined as the value of this observable after relaxation has occurred within the state $\t\rho_{x_0,t_0}$ describing the fluid cell. Relaxation occurs at mesoscopic time scales $t_{\rm meso}$; micro-, meso- and macroscopic timescales, in the cell $\t\rho_{x_0,t_0}$, are informally separated as
\begin{equation}\label{times}
    t_{\rm micro} \ll t_{\rm meso} \ll {\rm  min}_{l} ([v_{\rm micro} |\p_{x_0}\beta^l(x_0,t_0)|]^{-1}) \red {\equiv t_{\rm macro}}.
\end{equation}
Here the quantities $t_{\rm micro}, \, v_{\rm micro}$ are some microscopic time and velocity depending on the model. Averages of observables obtained after relaxation -- ``mesoscopic averages" -- are to be expressed as functions of the mesoscopic averages of conserved densities, which are then identified with the hydrodynamical variables ${\tt q}_i$. In practise, the mesoscopic averages are obtained by taking limits in the correct order: infinite macroscopic times $|\p_{x_0}\beta^l(x_0,t_0)|\sim 0$, followed by infinite microscopic evolution time $t\to\infty$.
\medskip

Utilising the hydrodynamic approximation, we can trivially write the conservation equations as
\beq\label{eq:hydroeqderivation}
\p_{t_0} {\tt q}_i(x_0,t_0) + \p_{x_0} \lim_{{\rm meso}}\bra j_{\boldsymbol w, i}(x_0,t_{\rm meso})\ket_{\t{\rho}_{x_0,t_0}} = 0,
\eeq
which is the basis of our hydrodynamic equation. Here $\lim_{{\rm meso}}$ is defined in eq. \eqref{times} and we have defined ${\tt q}_i(x_0,t_0)=\lim_{{\rm meso}}\bra q_i(x_0,t_{\rm meso})\ket_{\t{\rho}_{x_0,t_0}}$. The non-trivial task is to express the mesoscopic average of the currents in terms of ${\tt q}_i(x_0,t_0)$, which is the subject of the remainder of this section. Note that point II is crucial in establishing the irreversibility of the hydrodynamic equations based on \eqref{eq:hydroeqderivation}. In practise, the evolution over mesoscopic times at $x_0,t_0$ is obtained by linear response from the ramp state \eqref{eq:hydro1}: first performing a perturbation theory in $\p_{x_0} \beta^l(x_0,t_0)$, and then taking the infinite (microscopic) time limit. Thus the general procedure for obtaining the hydrodynamic equations is the following:
\beqa
    &&\bra o(x_0,t_0)\ket_{\rm ini} \n&\approx& \lim_{t\to\infty} \Big[\bra o(x_0,t) \ket_{\t{\rho}_{x_0,t_0}}\n && \hspace{1cm} \mbox{expressing $\beta^l(x_0,t_0)$ as functions of ${\tt q}_i^{(t)}(x_0,t_0) :=  \bra q_i(x_0,t)\ket_{\t{\rho}_{x_0,t_0}}$,}\n && \hspace{1cm} \mbox{expanding to first order in $\p_{x_0}\beta^l(x_0,t_0)$}
    \Big]\label{eq:method}
\eeqa
Under linear response from a ramp initial state, many observables grow linearly in time (such as currents of ballistically transported quantites). The fact that the long-time limit in the second line of \eqref{eq:method} exists and is finite as a function of ${\tt q}_i(x_0,t_0)$ is nontrivial, and must be ascertained by the explicit calculation.

Without loss of generality, it is sufficient to consider $x_0=t_0=0$. As written, \eqref{eq:hydro1} is sufficient to ascertain the diffusive hydrodynamics; neglecting the derivative term yields the Euler hydrodynamics, and including further terms in the series expansion of the argument of the exponential would yield higher derivative corrections to the hydrodynamic equation, which may or may not be physically sensible.

We denote $\t\rho = \t \rho_{0,0}$ and compute expectation values of charges $q_i(0,t)$ and their currents $j_{\ww,i}(0,t)$ in the state $\t \rho$; throughout we expand to first order in $\p_{x_0} \beta^l(x_0,0)|_{x_0=0} \equiv \partial_x \beta^l$. In the following, expectation values denoted $\langle \cdots \rangle$ are taken with respect to the homogeneous stationary GGE state $ {\rho}_{\rm GGE}  \propto e^{ -\beta^l(0,0) Q_l} $, in contrast with the case when they are taken with respect to the inhomogeneous ramp-state $\tilde{\rho}$  as $\langle \cdots \rangle_{\tilde{\rho}}$. 
We use the following relation, valid for generic operators $A$ and $B$
\begin{equation}
e^{A + \epsilon B } = e^A + \epsilon  \int_0^1 d\tau e^{\tau A}B e^{(1-\tau) A} + O(\epsilon^2)
\end{equation}
in order to expand the expectation values of charges and currents to first order in the derivatives, in terms of the KMB inner product \eqref{genKMB}. This gives, using the notation introduced in eq. \eqref{eq:evolwithw},
\begin{align}\label{eq:charges01}
 \langle q_i(0, t      \boldsymbol w   ) \rangle_{\tilde{\rho}} &= \langle  q_i \rangle  - \partial_x \beta^k K[q_k;q_i]  ,\\
\label{eq:currents01}
 \langle j_{\ww, i}(0, t      \boldsymbol w ) \rangle_{\tilde{\rho}} &= \langle  j_{\ww, i} \rangle  - \partial_x \beta^k   K[q_k;j_{ \ww,i }],
 \end{align}
 with the following expression for integrated correlation functions on a homogenous, stationary state:
\beqa
K[q_k;o] &=&    \int dy\, y  \,(  q_k(y )  , o(0, t \ww     ) )\nonumber \\
&=& \frac{1}{2}    \int dy\, y \int_{-t}^t ds\,\p_s  ( q_k(y ) ,  o(0, s \ww     ) ) \nonumber \\
&=& -\frac{1}{2}    \int dy \int_{-t}^t ds\, (  j_{\ww; k}(y ,s\ww)  , o(0,0     ) ), 
\label{eq:K}
\eeqa
where we used $PT$ symmetry, and stationarity of the state and the conservation equation, to introduce the $s$-integral, and current $j_{\ww,k}$, respectively.

As per \eqref{eq:method}, we now need to express $\beta^k$ in terms of ${\tt q}_i^{(t)} = \bra q_i(0,t)\ket_{\t\rho}$ and take the limit $t\to\infty$. This is obtained from \eqref{eq:charges01}, where the right hand side is considered as a function of $\beta^k$ and $\p_x\beta^k$,
and using \eqref{eq:K} we have:
\beq\label{eq:qiqi}
    \bra q_i\ket = {\tt q}^{(t)}_i +
    \p_x \beta^k K[q_k,q_i]
    = {\tt q}^{(t)}_i -
    t \,\p_x \beta^k (J_{\ww,k},q_i).
\eeq
Note that the second term on the right-hand side is small, as $t|\p_x \beta^k| \ll v_{\rm micro}^{-1}$. Consider Eq.~\eqref{eq:currents01}, where generally $\bra j_{{\ww,i}}\ket[\bra q\ket]$ is known from the thermodynamics, and thus we utilise this by changing variable to ${\tt q}_i^{(t)}$ within the same function. Eq.~\eqref{eq:qiqi} then implies:
\beq
\langle  j_{\ww,i} \rangle \equiv  \langle   j_{\ww,i} \rangle [\langle q \rangle] = \langle   j_{\ww,i} \rangle[{\tt q}^{(t)}] - t\,\partial_x \beta^k  \frac{\delta \langle j_{\ww,i} \rangle}{\delta \langle q_l \rangle} (J_{\ww,k},q_l) + \ldots,
\eeq
and therefore \eqref{eq:currents01} can be expressed as:
\beqa\label{eq:jwitw}
    \bra j_{\ww,i}(0,t      \boldsymbol w )\ket_{\t\rho} &=&
    {\tt j}_{\ww,i} + 
    \frc{\p_x \beta^k}2 \int dy \int_{-t}^t ds\,
     (j_{\ww,k}(y,s\ww),j_{\ww, i}(0,0) -
    \frac{\delta \langle j_{\ww,i} \rangle}{\delta \langle q_l \rangle}q_l(0,0)) \n &=&
    {\tt j}_{\ww,i} + 
    \frc{\p_x \beta^k}2 \int dy \int_{-t}^t ds\,
    (j_{\ww,k}(y,s\ww),j_{\ww, i})^C.
\eeqa
where in the first line we used the definition \eqref{eq:jGGE}, in the second, the chain rule for differentiation and the projection formula \eqref{eq:projection}. Taking the limit $t\to\infty$, this returns the hydrodynamic equation in the absence of external forces:
\begin{equation}
    \p_t{\tt q}_i+\p_x\left({\tt j}_{\ww,i} + \frac{1}{2} \mathfrak L[\ww,\p_x \boldsymbol\beta;j_{\ww,i}]\right)=0
\end{equation}
Two subtleties need to be clarified in the obtaining of this hydrodynamic equation in the final step.

First, in \eqref{eq:jwitw}, the Euler-scale current ${\tt j}_{w,i}$ is evaluated within a GGE  characterised by the ${\tt q}_i$'s, while the KMB inner product $(\cdot,\cdot)$ is evaluated within the GGE determined by $\boldsymbol\beta(0,0)$. These are {\em different GGE's}, expectation values of charge densities within the GGE with potentials $\boldsymbol\beta(0,0)$ are $\bra q_i\ket\ne{\tt q}_i$; the correction term on the right-hand side of \eqref{eq:qiqi} means that ${\tt q}_i$ can be written as a GGE average of charge densities by the expected bijectivity between potentials and charge densities, but with different associated thermodynamic potentials as per \eqref{eq:hydrostate}, say $\boldsymbol\beta^{\rm hydro}(0,0)$. But the difference is first-order in derivatives, and hence in \eqref{eq:jwitw} we can use $\boldsymbol\beta^{\rm hydro}(0,0)$ for the KMB inner product, the error being second-order in derivatives.

Second, in \eqref{eq:jwitw}, $\p_x\boldsymbol\beta$ is the slope of the ramp in the density matrix $\t\rho$. We must equate the slope $\p_x\boldsymbol\beta$ with the corresponding macroscopic spatial derivative $\p_{x_0}\boldsymbol\beta^{\rm hydro}(x_0,0)|_{x_0=0}$, connecting neighbouring fluid cells. This can be done using \eqref{eq:qiqi}, written for arbitrary $x_0$, and taking the derivative $\p_{x_0}$. As $t\p_x \beta^k\ll v_{\rm micro}^{-1}$, the correction term is one derivative order smaller, and thus again to leading order $\p_x\boldsymbol\beta = \p_{x_0}\boldsymbol\beta^{\rm hydro}(x_0,0)|_{x_0=0}$, which is sufficient in \eqref{eq:jwitw}. With this identification, the results of section \ref{sec:intro} are technically expressed in terms of $\boldsymbol\beta^{\rm hydro}$ rather than the potentials introduced in \eqref{eq:density_mat}, as these form the most convenient basis to describe the evolution.

\subsection{Diffusive hydrodynamics: inhomogeneous fields }
We now turn our attention again to spatially modulated fields $w^i(x)$ in the Hamiltonian \eqref{eq:Hamiltonian}. As we did for the state above, we can also expand the Hamiltonian locally around any point $x_0$ as
\begin{equation}\label{eq:pert}
H= w^i(x_0) Q_i -  \mathfrak{f}^i(x_0) \int dx\, (x-x_0)   q_i(x) + \ldots,
\end{equation}
providing similar conditions hold. As we are interested in a hydrodynamic theory up to second derivatives, it is in principle necessary to also include the second order term this expression. However imposing PT symmetry for all densities $q_i(x)$ leads such contributions to the hydrodynamic equations to vanish, see appendix \ref{app:PT}. 

The continuity equation for the charge and current density operators now has an extra contribution, due to the new term in the Hamiltonian, and reads
\begin{equation}\label{eq:newCE}
\partial_t q_i( x,t) = - \partial_x j_{\ww,i} (x,t  ) +  j_{i,\boldsymbol{\mathfrak{f}}}(x,t)  + \ldots,
\end{equation}
where we have used the result \cite{SciPostPhys.2.2.014} (a higher-dimensional generalisation is shown in \cite{Doyon2021}):
\begin{equation}
    \ii \left[q_k(y),q_i(x)\right]=\p_xj_{k,i}(x)\delta(y-x)+(j_{k,i}(x)+j_{i,k}(x))\delta'(y-x)+\left(\mathrm{higher}\;\mathrm{derivatives}\right).\label{commutatorqq}
\end{equation}

It should be noted that now $[ H, Q_i] \ne 0$, but, as mentioned, in the hydrodynamic framework, the full set of charges for the homogeneous part of \eqref{eq:pert} is to be considered, as the correction is most aptly treated through a separation of scales. This means that we assume that the typical scale of spatial modulation of the external  fields, $\ell_{\mathfrak f} = {\rm min}_i (|\mathfrak f^i|^{-1})$, must be large enough. As mentioned, $\ell_{\mathfrak f}$ naturally determines the scale of spatial modulation of the local fluid state, hence we expect to have, at all large enough times, $\ell\approx\ell_{\mathfrak f}$, and the result of the hydrodynamic framework is an equation valid up to, and including, order $\ell^{-2}\approx \ell_{\mathfrak f}^{-2}$. The physical interpretation is that the correction generates a fast dynamics to a local hydrodynamic state as eq. \eqref{eq:hydrostate} on mesoscopic times, and the effect on the state due to the perturbations can be evaluated by perturbation theory in the interaction picture. Again, as in our discussion of the case without external forces, taking the long-time limit in the perturbation theory amounts to taking a mesoscopic time scale and accounting for local relaxation,
\begin{equation}\label{timesforce}
t_{\rm micro}\ll t_{\rm meso} \ll {\rm min}_{l,x_0} ((v_{\rm micro}|\p_{x_0}\beta^l(x_0,t_0)|)^{-1},\,(v_{\rm micro}|\mathfrak f^l(x_0)|)^{-1}).
\end{equation}
Once the average currents have been obtained in fluid cells after mesoscopic times, we insert them into \eqref{eq:newCE} in order to obtain the final hydrodynamic equation.

We employ perturbation theory to first order in the perturbation strength, $ \mathfrak{f}^i(0)$, as higher orders contain higher spatial derivatives and powers thereof. Under the full time evolution generated by $H$ eq. \eqref{eq:pert}, we find, for a generic local operator $o(x,t) = e^{i H t} o(x) e^{-i H t}$ at position $x=0$, 
\begin{equation}
o(0,t) = o(0,t \ww) -   \ii   \int_0^t d s\, \int dx \ x \ \mathfrak{f}^i(0) \  [q_i(x,s \ww),o(0,t \ww)]  +\ldots,
\end{equation}
where, as before, we denoted operators evolved with respect to the portion of the Hamiltonian with flat fields in eq. \eqref{eq:pert} by means of the time arguments $t \ww$, as in eq. \eqref{eq:evolwithw}. A single time argument refers still to the real time under the full evolution. The hydrodynamic expansion of the previous section in the presence of these new terms now reads
\begin{align}\label{eq:q_pretherm}
 & \langle q_i(0,t) \rangle_{\tilde{\rho}}  =  \langle  q_i \rangle - (\partial_x \beta^k) K[q_k;q_i]  -   \mathfrak{f}^k \overline{K}[q_k; q_i],
\end{align} 
\begin{align}\label{eq:j1_pretherm}
 & \langle j_{\ww,i}(0,t) \rangle_{\tilde{\rho}}  = \langle  j_{\ww,i} \rangle - (\partial_x \beta^k) K[q_k ;  j_{\ww,i}]  -    \mathfrak{f}^k   \overline{K}[q_k; j_{\ww,i}],
\end{align} 
\begin{align}\label{eq:j2_pretherm}
 & \langle j_{i,\boldsymbol{\mathfrak{f}}}(0,t) \rangle_{\tilde{\rho}}  = \langle j_{i,\boldsymbol{\mathfrak{f}}} \rangle - (\partial_x \beta^k) K[q_k ;j_{i,\boldsymbol{\mathfrak{f}}}] -    \mathfrak{f}^k   \overline{K}[q_k;j_{i,\boldsymbol{\mathfrak{f}}}],
\end{align} 
where $K[a;b]$ is defined in \eqref{eq:K}, and we define the integrated correlation functions of the commutators as 
\begin{equation}\label{eq:KKs}
  \overline{K}[a;b] = \ii  \int_0^t d s\, \int dy \,y \,\langle  [a(y,s \ww),b(0,t \ww)]   \rangle.
\end{equation}

We now invoke hydrodynamic separation of scales as in the previous section with homogeneous Hamiltonians, assuming that the space-time dependence of the system is contained entirely in the potentials $\beta^k(x,t)$, which vary slowly in space and time. Here, this consists of taking the infinite time limit of Eqs.~\eqref{eq:q_pretherm} - \eqref{eq:j2_pretherm}, which describe the slow drift in the values of the charges after relaxation under the homogenous-breaking perturbation has occurred, where we again assume that the limit is approximately reached at timescales much shorter than the timescales of hydrodynamic evolution. Proceeding analogously to the previous section, and taking the infinite time limit in eq. \eqref{eq:KKs} we obtain the following diffusive equation
\begin{align}\label{eq:finalperturb}
\partial_t {\tt q}_i &+\partial_x \Big[
 {\tt j}_{\ww,i}    + \frac{1}{2} \mathfrak{L}[ \ww , \partial_x\boldsymbol{\beta}; j_{\ww,i}] 
	-  \frac{1}{2} \mathfrak{F}[ \boldsymbol{\mathfrak{f}} , j_{\ww,i}]  \Big] 
	=  { \tt j}_{i, \boldsymbol{\mathfrak{f}}}   +  \frac{1}{2} \mathfrak{L}[ \ww , \partial_x\boldsymbol{\beta} ;j_{i;\boldsymbol{\mathfrak{f}}}] 
	-  \frac{1}{2} \mathfrak{F}[  \boldsymbol{\mathfrak{f}} ,j_{i;\boldsymbol{\mathfrak{f}}}]  ,
\end{align}
with the new integrated correlator defined as by
\begin{align}\label{deffrakF}
& \mathfrak{F}[  \boldsymbol{\mathfrak{f}} ,o] = {\ii}   \int_{-\infty}^{\infty} ds    \int dy \ y   \  \mathfrak{f}^k \langle    [q_{k}(y, \boldsymbol{w} s ) ,    o(0,0) - \frc{\p \bra o\ket}{\p \bra q_i\ket} q_i(0,0)] \rangle ,
\end{align}
where again we have used PT symmetry in order to symmetrise the integral over time. Here, the state $\bra\cdots\ket$ is the GGE at the fluid cell $(x,t)$.

The latter quantity can be rewritten in terms of the usual Onsager coefficients by employing the Kubo–Martin–Schwinger (KMS) relation \cite{PhysRev.115.1342,Doyon2021}. This allows us to rewrite the expectation value of the commutator in any homogeneous, stationary GGE as
\begin{align}
    \langle [q_k(y,s \ww), o(0,0)] \rangle =& - \ii \int_0^1 d\lambda \,\beta^l \partial_y \langle j_{l,k}(y,s \ww- \ii \lambda \boldsymbol{\beta}) o(0,0)  \rangle
    \nonumber \\=& -\ii\, \p_y (j_{\boldsymbol\beta,k}(y,s \ww),o(0,0)).
\end{align}
Note that the spatial derivative along with homogeneity of the state allows us to introduce the connected correlation function.
Therefore we have, integrating by parts over $y$ in \eqref{deffrakF},
\begin{align}
 \mathfrak{F} [  \boldsymbol{\mathfrak{f}} ,o] = 
 -\mathfrak{L}[\boldsymbol{\beta}, \boldsymbol{\mathfrak{f}} ; o],
\end{align}
which finally gives our main result \eqref{main_result}.
\section{Integrable systems: quasiparticle expression and lower bounds for Onsager coefficient}\label{sec:int_systems}

In integrable systems, the local GGE state is most conveniently characterised by the Bethe rapidities $\theta$, whose distribution function (root density) in the fluid cell located at $(x,t)$ is denoted $\rho_{\rm p }(\theta; x,t)$. This object specifies the density of quasiparticles with rapidity $\theta$ at position $(x,t)$, and is related to the hydrodynamic variables by
\begin{equation}
{\tt  q}_i(x,t)  = \int \ d \theta  \rho_{\rm p }(\theta;x,t) h_i(\theta),
\end{equation}
where $h_i(\theta)$ are the single-particle eigenvalues of the charges $Q_i$. With a sufficiently complete set of charges, it is possible to invert this expression, so there is a bijection between the root density and the expectation values of the charges. We now introduce the dressing operation on functions defined over $\mathbb{R}$. The action of a linear integral operator is:
\begin{equation}
    A\cdot h:=\int d\alpha A(\theta,\alpha)h(\alpha), \quad 
    \quad  h \cdot A:=\int d\alpha h(\alpha) A(\alpha,\theta),
\end{equation}
with which we can define the dressed function
\begin{equation}
h^{\rm dr} = (1- T n)^{-1} \cdot h,
\end{equation}
where $[Tn](\theta,\alpha)=T(\theta,\alpha)n(\alpha)$. Here $T$ is the scattering shift of the model, which is independent of the state and taken to be symmetric $T(\theta,\alpha)=T(\alpha,\theta)$. The filling function is $n = 2\pi \rho_{\rm p }/(p')^{\rm dr}$, where $p(\theta)$ is the eigenvalue of the momentum operator. In the following we shall denote $(p')^{\rm dr} = k'$.

The main task is to compute the hitherto unknown expressions for the extended Onsager matrices $\mathfrak{L}[\boldsymbol{a},  \boldsymbol{b} ; j_{\boldsymbol{c},\boldsymbol{d}}]$ defined by \eqref{eq:def_Onsager}. In integrable models these can be computed exactly whenever  { at least one of the currents $j_{\boldsymbol a,\boldsymbol b}$ or $j_{\boldsymbol c,\boldsymbol d}$ project entirely on the space spanned by linear and quadratic fluctuations of the local conserved densities \cite{10.21468/SciPostPhys.9.5.075,1912.01551}. This is argued to be the case if $\boldsymbol a = \boldsymbol w$ or $\boldsymbol c = \boldsymbol w$, as, after hydrodynamic reduction, currents of the type $j_{\boldsymbol w,\boldsymbol b}$ lie in the hydrodynamic subspace that is invariant under higher flows, which is argued to be spanned by quadratic charges \cite{Durnin2020,1912.01551}.}
In this case, the only contributions to diffusion are given by two-body scattering amongst quasiparticles: the two particle-hole contribution in the expansion over intermediate states $\mathfrak{L}[\boldsymbol{a},  \boldsymbol{b} ; j_{\boldsymbol{c},\boldsymbol{d}}]= \mathfrak{L}_{2-\rm ph}[\boldsymbol{a},  \boldsymbol{b} ; j_{\boldsymbol{c},\boldsymbol{d}}] $ \cite{PhysRevLett.121.160603,1911.01995,PhysRevB.98.220303,RevCorrelations}. The idea that such two-body processes are responsible for diffusion in integrable models first appeared before the advent of GHD in \cite{PhysRevLett.83.2293}. We compute the two particle-hole contribution analogously to Ref.~\cite{10.21468/SciPostPhys.6.4.049} via a form-factor expansion, see appendix \ref{sec:FFComputation}.

{ However, currents of the type $j_{\boldsymbol a,\boldsymbol b}$ for $\boldsymbol a\neq \boldsymbol w$ do not entirely overlap with the space of linear and quadratic fluctuations. Therefore, while three of the four Onsager coefficients in \eqref{main_result} are fully given by their two particle-hole contributions, the coefficient $\mathfrak{L}[\boldsymbol{  \beta},\boldsymbol{\mathfrak{f}}  ; j_{i,\boldsymbol{\mathfrak{f}}}]$ is not, see appendix \ref{sec:FFComputation} for more details. Nevertheless, the lower bound $\mathfrak L\geq \mathfrak{L}_{2-\rm ph}$ for the extended Onsager matrix \eqref{OnsagerMatrix}
follows from the hydrodynamic projection mechanism and the identification of the two particle-hole contribution with the projection onto the quadratic space \cite{10.21468/SciPostPhys.9.5.075,1912.01551}.}

Our results read
\beqa\label{Lintegrable1}
    \mathfrak{L}[\boldsymbol{w}, \partial_x\boldsymbol{\beta} ; j_{\boldsymbol{w},i}] &=&   h_i \cdot  \mathfrak{D} \mathsf C \cdot  \partial_x( \beta^k h_k  ),\\
    \mathfrak{L}[\boldsymbol{  \beta},\boldsymbol{\mathfrak{f}}  ; j_{\boldsymbol{w},i}] &=&   h_i \cdot \mathfrak{D}_{\boldsymbol{\mathfrak{f}}} \mathsf C  \cdot     \partial_\theta( \beta^k h_k),\\
    \mathfrak{L}[\boldsymbol{w}, \partial_x\boldsymbol{\beta} ; j_{i,\boldsymbol{\mathfrak{f}}}] &=&         \partial_\theta h_i   \cdot \mathfrak{D}_{\boldsymbol{\mathfrak{f}}} \mathsf C \cdot   \partial_x (\beta^k  h_k) ,\\
    \mathfrak{L}[\boldsymbol{  \beta},\boldsymbol{\mathfrak{f}}  ; j_{i,\boldsymbol{\mathfrak{f}}}]  & = &             \partial_\theta h_i \cdot \mathfrak{D}_{\boldsymbol{\mathfrak{f}^2}} \mathsf C \cdot  \partial_\theta( \beta^kh_k ) .
    \label{Lintegrable4}
\eeqa
The kernels in these equations are defined below. All kernels can be written in terms of the full effective velocity given by the dynamics of the system Hamiltonian \eqref{eq:Hamiltonian} at position $x$, namely 
\begin{equation}
v^{\rm eff}_{\ww}(\theta; x,  t )= \frac{ w^i(x)  (h_i')^{\rm dr}(\theta; x ,t )}{k'(\theta; x , t )},
\end{equation}
and of the effective acceleration \cite{SciPostPhys.2.2.014},
\beq
    a^{\rm eff}_{\boldsymbol{\mathfrak f}}(\theta; x ,t ) = \frac{ \mathfrak f^i(x)  (h_i)^{\rm dr}(\theta; x ,t )}{k'(\theta; x , t )}.
\eeq
The kernels in rapidity space are given by
\begin{align}\label{eq:DD}
 \mathfrak{D} \mathsf C   = (1- n T )^{-1}\cdot \frac{\delta_{\theta_1,\theta_2} \int d\alpha  {\kappa_{\mathfrak{D}}(\theta_1,\alpha)}{ } -  {\kappa_{\mathfrak{D}}(\theta_1,\theta_2)}{} }{k'(\theta_1) k'(\theta_2)}   \cdot  (1-   T n  )^{-1},
\end{align}
and similarly for $\mathfrak D_{\boldsymbol{\mathfrak f}},\,\mathfrak D_{\boldsymbol{\mathfrak f}^2}$ with different functions $\kappa_{\mathfrak D_{\boldsymbol{\mathfrak f}}},\,\kappa_{\mathfrak D_{\boldsymbol{\mathfrak f}^2}}$,
where the susceptibility kernel is given by 
\begin{equation}
    \mathsf C=  (1- n T )^{-1} \cdot \rho_{\rm p} f \cdot (1-  T n )^{-1},
\end{equation}
with the function $f(\theta)$ incorporating the statistics of quasiparticles (see for example the review \cite{RevCorrelations}). Notice that $ \rho_{\rm p} f$ or its inverse denote the diagonal operator $\rho_{\rm p}(\theta) f(\theta) \delta(\theta - \alpha)$.
The function $\kappa_{\mathfrak{D}}$ is defined as (neglecting the $x,t$ dependence)
\begin{align}\label{eq:ffun}
\kappa_{\mathfrak{D}}(\theta_1,\theta_2) & = k'(\theta_1) k'(\theta_2)     n(\theta_1) f(\theta_1) n(\theta_2) f(\theta_2) (T^{\rm dr}(\theta_1,\theta_2))^2 |v^{\rm eff}_{\ww}(\theta_1) - v^{\rm eff}_{\ww}(\theta_2)|  ,
\end{align}
and the others are defined by including ratios of effective velocity and acceleration differences
\beqa
\kappa_{\mathfrak D_{\mathfrak f}}(\theta_1,\theta_2) &=& \frac{a^{\rm eff}_{\boldsymbol{\mathfrak f}}(\theta_1)-a^{\rm eff}_{\boldsymbol{\mathfrak f}}(\theta_2)}{v^{\rm eff}_{\ww}(\theta_1) - v^{\rm eff}_{\ww}(\theta_2)}\times \kappa_{\mathfrak{D}}(\theta_1,\theta_2) \label{eq:kdf}\\
\kappa_{\mathfrak D_{\mathfrak{ f}^2}}(\theta_1,\theta_2) &\stackrel{{2-\rm ph}}=& \lt(\frac{a^{\rm eff}_{\boldsymbol{\mathfrak f}}(\theta_1)-a^{\rm eff}_{\boldsymbol{\mathfrak f}}(\theta_2)}{v^{\rm eff}_{\ww}(\theta_1) - v^{\rm eff}_{\ww}(\theta_2)}\rt)^2\times \kappa_{\mathfrak{D}}(\theta_1,\theta_2) \label{eq:kdf2}.
\eeqa

The general multi-linear expression \eqref{Labcd} for the Onsager coefficient makes it clear that in thermal states \eqref{LDA_state}, stationarity occurs, with the following cancellation:
\begin{align}
    & \mathfrak{L}^{\rm thermal}[\boldsymbol{w}, \partial_x\boldsymbol{\beta} ; j_{\boldsymbol{w},i}]+  \mathfrak{L}^{\rm thermal}[\boldsymbol{  \beta},\boldsymbol{\mathfrak{f}}  ; j_{\boldsymbol{w},i}]=0, \\
     & \mathfrak{L}^{\rm thermal}[\boldsymbol{w}, \partial_x\boldsymbol{\beta} ; j_{i,\boldsymbol{\mathfrak{f}}}] + \mathfrak{L}_{2-\rm ph}^{\rm thermal}[\boldsymbol{  \beta},\boldsymbol{\mathfrak{f}}  ; j_{i,\boldsymbol{\mathfrak{f}}}]  =0.
\end{align}
Ultra-local quantities, which have an arbitrary chemical potential in the thermal states, correspond in the quasi-particle basis to quantities $Q_0$ for which $h_0(\theta)=$ a constant. It is possible to use the following relations 
\begin{equation}\label{eq:thermal1}
    \partial_x( \beta^i h_i) = - (1- T n) \cdot \frac{\partial_x n}{ n f(n)} ,
\end{equation}
\begin{equation}\label{eq:thermal2}
    \partial_\theta( \beta^i h_i) = - (1- T n) \cdot\frac{\partial_\theta n}{ n f(n)} .
\end{equation}
to verify that expressions \eqref{Lintegrable1}-\eqref{Lintegrable4} do satisfy the above cancellations in a thermal state: In thermal states, the occupation number satisfies the following relations:
\begin{align}\label{eq:thermalpp}
   &  \partial_\theta n^{\rm thermal}  = - \beta n  f  v^{\rm eff}_{\ww}  k' , \quad \quad
     \partial_x n^{\rm thermal}   =  \beta n   f a^{\rm eff}_{\boldsymbol{\mathfrak{f}}}  k'
\end{align}
which, together with $\partial_x \beta^i = -\beta \mathfrak{f}^i$ in the thermal state, implies the above cancellation between the sum of the Onsager coefficients.

We are now in a position to write Eq.~\eqref{main_result} for quasiparticles, using the relation $\delta n = \frac{n}{\rho_{\rm p}} (1- n T) \delta \rho_{\rm p}$, where $\delta$ is a variation with respect to the rapidity or the spatial argument. We use this to obtain the following hydrodynamic equation
\begin{align}\label{main_particle_result}
\partial_t \rho_{\rm p }     =    \begin{pmatrix} \partial_x   & \partial_\theta  \end{pmatrix} \cdot \Big[ \frac{1}{2} \begin{pmatrix}   \mathfrak{D} &   \mathfrak{D}_{\boldsymbol{\mathfrak{f}}}  \\  \mathfrak{D}_{\boldsymbol{\mathfrak{f}}}  &   \mathfrak{D}_{\boldsymbol{\mathfrak{f}^2}}  \end{pmatrix}    \cdot \begin{pmatrix}  \partial_x \rho_{\rm p}  \\ \partial_\theta \rho_{\rm p} \end{pmatrix} - \begin{pmatrix} v^{\rm eff}_{\ww}  \rho_{\rm p }  \\ a_{\boldsymbol{\mathfrak{f}}}^{\rm eff} \rho_{\rm p } \end{pmatrix} \Big],
\end{align}
where we have used the known results for the Euler currents
\cite{Bertini16,PhysRevX.6.041065,SciPostPhys.2.2.014,PhysRevX.10.011054,10.21468/SciPostPhys.8.2.016,2004.07113}
\begin{align}
 & { \tt j}_{\ww, i} (x,t )= \int  d\theta \ \rho_{\rm p}(\theta; x, t) v_{\ww}^{\rm eff}(\theta; x ,t ) h_i(\theta), \\
    & { \tt j}_{i, \boldsymbol{\mathfrak{f}}}(x,t) = \int  d\theta \ \rho_{\rm p}(\theta; x ,t ) a_{\boldsymbol{\mathfrak{f}}}^{\rm eff}(\theta; x ,t ) h_i'(\theta).
\end{align}
A notable consequence of eq. \eqref{main_particle_result} is that motion of the fluid due to inhomogeneities in the state always corresponds to spatial derivatives, while the effect of the external forces is contained in rapidity derivatives. Therefore eq. \eqref{main_particle_result} describes convective and diffusive motions for the quasiparticles, equally in the physical and quasi-momentum coordinate of the fluid. In particular the kernel $ \mathfrak{D}_{\boldsymbol{\mathfrak{f}^2}}$ can be understood as the effective diffusion constant of the quasiparticles in momentum space. The presence of the latter is directly an effect of breaking the underlying integrability of the system, by means of the forces $\mathfrak{f}^i$, and proportional to the square of the force strength, similarly to Fermi's golden rule terms in homogeneous settings \cite{Durnin2020,2103.11997,2005.13546}. We stress that our analytical expression for $\mathfrak{D}_{\boldsymbol{\mathfrak{f}^2}}$, given in terms of the function \eqref{eq:kdf2}, constitutes generically only a lower bound for such effective diffusion constant, where only two-body scatterings processes are taken into account. Higher particle scattering processes,  can indeed non-trivially contribute to this kernel, { nevertheless the form \eqref{Lintegrable4} is expected to hold at all orders.}

As the analysis here is conducted for integrable systems with only one quasiparticle type, the extension to systems with multiple species is easily done by replacing the rapidity $\theta$ with the global index $(\theta,s)$, and inserting a sum over particle species to accompany each rapidity integral.

It is easy to check that the total density of quasiparticles $N$ and the total energy $E = \langle H \rangle$ are conserved quantities of motion. We have
\begin{equation}
  \partial_t N =  \int d\theta \int dx  \ \partial_t \rho_{\rm p}(\theta;x,t) =0
\end{equation}
by the structure of derivatives in the equation. Regarding the total energy
\begin{equation}
\partial_t E = \int d\theta \int dx  \ h_i(\theta) w^i(x)\  \partial_t \rho_{\rm p}(\theta;x,t)    
\end{equation}
it is already established that the convective terms $\partial_x \left( v^{\rm eff}_{\ww}  \rho_{\rm p }\right) +     \partial_\theta  \left(   a_{\boldsymbol{\mathfrak{f}}}^{\rm eff} \rho_{\rm p } \right)$ in eq. \eqref{main_particle_result} conserve total energy, \cite{SciPostPhys.2.2.014} while for the diffusive terms we have, after integration by parts
\begin{align}
    \partial_t E &  =\frac{1}{2} \int dx \int d\theta \ \Big[   \mathfrak{f}^i h_i \cdot \mathfrak{D} \cdot \partial_x \rho_{\rm p } +      \mathfrak{f}^i h_i  \cdot  \mathfrak{D}_{\boldsymbol{\mathfrak{f}}} \cdot \partial_\theta \rho_{\rm p }     
\nonumber \\& -  w^i h_i'    \cdot  \mathfrak{D}_{\boldsymbol{\mathfrak{f}}} \cdot \partial_x \rho_{\rm p } - w^i h'_i \cdot  \mathfrak{D}_{\boldsymbol{\mathfrak{f}^2}} \cdot \partial_\theta \rho_{\rm p}     \Big]=0,
\end{align}
as by their definitions, eq. \eqref{eq:DD} together with eq. \eqref{eq:kdf}, \eqref{eq:kdf2},  we have 
\begin{align}
       & \mathfrak{f}^i h_i \cdot \mathfrak{D}  = w^i h_i' \cdot    \mathfrak{D}_{\boldsymbol{\mathfrak{f}}} , \\
       & \mathfrak{f}^i h_i  \cdot  \mathfrak{D}_{\boldsymbol{\mathfrak{f}}}  = w^i h'_i \cdot  \mathfrak{D}_{\boldsymbol{\mathfrak{f}^2}}. \label{eq:energyconsconstraint}
\end{align}
The positive entropy increase \eqref{eq:entropyIncrease} can also be written explicitly in the quasiparticle basis by projecting it into the two particle-hole sector. Denoting $\Delta_{x} n=  (1- T n) \cdot \frac{\partial_x n}{ n f(n)} $ and $\Delta_{\theta} n= (1- T n) \cdot \frac{\partial_\theta n}{ n f(n)}$, the entropy increase can be written as a quadratic $2$-by-$2$ block quadratic form:  
\begin{align}\label{eq:entropyLower}
    \partial_t S &  = \frac{1}{2} \begin{pmatrix} \Delta_x n & \Delta_\theta n \end{pmatrix} \cdot \begin{pmatrix}   \mathfrak{D}\mathsf C &   \mathfrak{D}_{\boldsymbol{\mathfrak{f}}}\mathsf C  \\  \mathfrak{D}_{\boldsymbol{\mathfrak{f}}} \mathsf C &   \mathfrak{D}_{\boldsymbol{\mathfrak{f}^2}} \mathsf C \end{pmatrix}    \cdot \begin{pmatrix} \Delta_x n \\ \Delta_\theta n \end{pmatrix} .
\end{align}
{ Note that both $ \mathfrak{D}\mathsf C$ and $\mathfrak{D}_{\boldsymbol{\mathfrak{f}^2}} \mathsf C$ are positive semi-definite operators, as is the full matrix of operators on the right-hand side of \eqref{eq:entropyLower}.} As our explicit expression for $\mathfrak{D}_{\boldsymbol{\mathfrak{f}^2}}\mathsf C$ is a lower bound, substituting it into eq. \eqref{eq:entropyLower} provides a lower bound for entropy production. However, such a lower bound still ensures positive entropy production and vanishing of entropy generation on thermal states, as it can be easily verified using eq. \eqref{eq:thermalpp}.

\section{Thermalisation in the Toda gas}\label{sec:toda}
\subsection{Thermodynamics}
The most immediate consequence of the formalism derived in the previous sections, is that integrable systems perturbed by couplings to the local charge densities thermalise in the diffusive regime. This is a prediction accessible by molecular dynamics simulations of classical systems. We choose as an example the integrable Toda system, whose Hamiltonian in the presence of an external field coupling to the energy can be written
\begin{equation}\label{Toda_Ham}
    H=\sum_{i=1}^N\frac{V(x_i)}{2}\left(p_i^2+e^{-(x_{i+1}-x_i)}+e^{-(x_i-x_{i-1})}\right),
\end{equation}
where we have absorbed the effect of the external potential into a single prefactor $V(x_i)$. In fact, in the integrable Toda system there are two complementary interpretations, whose notions of space differ. In the first case, which we consider here, the Hamiltonian \eqref{Toda_Ham} is viewed as describing a gas of particles with position $x_i$, which is however not invariant under permutations of these coordinates. In this case the physical space is parameterised by the $x_i$, and the choice of external field in \eqref{Toda_Ham} reflects our consideration of the gas picture. In the complementary chain picture, the physical space is parameterized by the index $i$, and we would have instead a potential $V_i$. Nevertheless, the two systems are expected to be thermodynamically equivalent, and while the dynamics of the two systems will differ in the presence of external fields, we expect that our general results will apply to both, with the suitable definition of physical space. See \cite{Doyon2019} for a detailed discussion of the relationship between the gas and chain pictures. 

We will take periodic boundary conditions $x_{N+1}=x_1+L$ and $x_{0}=x_N-L$; in the absence of the external field such that $V(x_i)=1$, the system with these boundary conditions is integrable. The full thermodynamics of the integrable system were recently elucidated in \cite{doyon_generalised_2019,spohn_generalized_2019} for the open system, following earlier analysis of the thermal states (see the review \cite{cuccoli_thermodynamics_1994} and references therein). The dynamics of the open system are unbounded, however one can bound the dynamics by introducing a pressure term, which under the equivalence of ensembles is expected to provide equivalent results to the periodic system. We define the variable $R=x_N-x_1$ which is approximately $L$ in the periodic case due to energetic considerations, as long as $L\sim N$, which we shall impose. The dynamics of the integrable Toda system in various contexts have recently been studied both analytically within the context of Euler-GHD and numerically \cite{Bulchandani_2019,spohn_ballistic_2020,spohn_hydrodynamic_2021,mendl_high-low_2021}.

In order to make the connection between the numerical results and the analytical predictions, we define numerical `fluid cells' of width $w\gg L/N$, and compare the average values of observables within these cells in the stationary state to those in the state defined by the maximal entropy condition \eqref{eq:conditionSmax}. This process is simplified by the fact that, in the thermal state, the thermodynamics of the Toda gas can be solved explicitly. In the Toda gas, there are two convenient ensembles which can be used, the Gibbs ensemble in which $N$ is fixed and $R$ is fluctuating, and the Landau ensemble in which $R$ is fixed and $N$ fluctuates. In the simulation procedure, the bins are of fixed width, and therefore the latter ensemble is the relevant one, although it is convenient to first evaluate the former, in order to express the thermodynamics in the latter. Introducing the pressure $P$ and the chemical potential $\mu$, we have
\begin{align}
    \mathcal{Z}_\mathrm{Gibbs}=&\int \prod_{i=1}^Ndx_idp_i\exp\left(\frac{-\beta p_i^2}{2}\right)\exp\left(-\beta e^{-(x_{i+1}-x_i)}-P(x_{i+1}-x_i)\right)\sim e^{-Ng},
\end{align}
\begin{align}
    \mathcal{Z}_\mathrm{Landau}=&\sum_{N=1}^\infty e^{\mu N}\int \prod_{i=1}^Ndx_idp_i\exp\left(\frac{-\beta p_i^2}{2}\right)\exp\left(-\beta e^{-(x_{i+1}-x_i)}\right)\delta(x_N-x_1-R)\sim e^{-Rf}.
\end{align}
The thermodynamics in the Gibbs ensemble is found to be
\begin{equation}\label{toda_therm}
    \varep(P,\beta)=\frac{\exv{H}}{N}=\frac{2P+1}{2\beta}\;,\;\;\nu(P,\beta)=\frac{\exv{R}}{N}=\log(\beta)-\p_P\Gamma(P).
\end{equation}
We relate the two ensembles using \cite{Doyon2019}:
\begin{equation}
    f(\mu,\beta)+P=\frac{g(P,\beta)-\mu}{\nu},
\end{equation}
where the ensembles are chosen such that the density $\nu^{-1}$ is the same in both, where in the Landau ensemble $\nu=R/\exv{N}$. This allows the calculation of expectation values in the Landau ensemble, and thus to compare the LDA state predictions against the numerical results. There is a subtlety when comparing with the numerics, arising from the fact that $R$ is not necessarily positive under the dynamics, or in the thermodynamics. If this occurs, then matching the thermodynamics as calculated through bins of fixed width becomes ill-defined. Therefore the initial state is chosen such that $R>0$ for every sufficiently large subset of particles in the gas for all times.
\subsection{Numerical Results}
\begin{figure}
    \centering
    \includegraphics[width=0.485\textwidth]{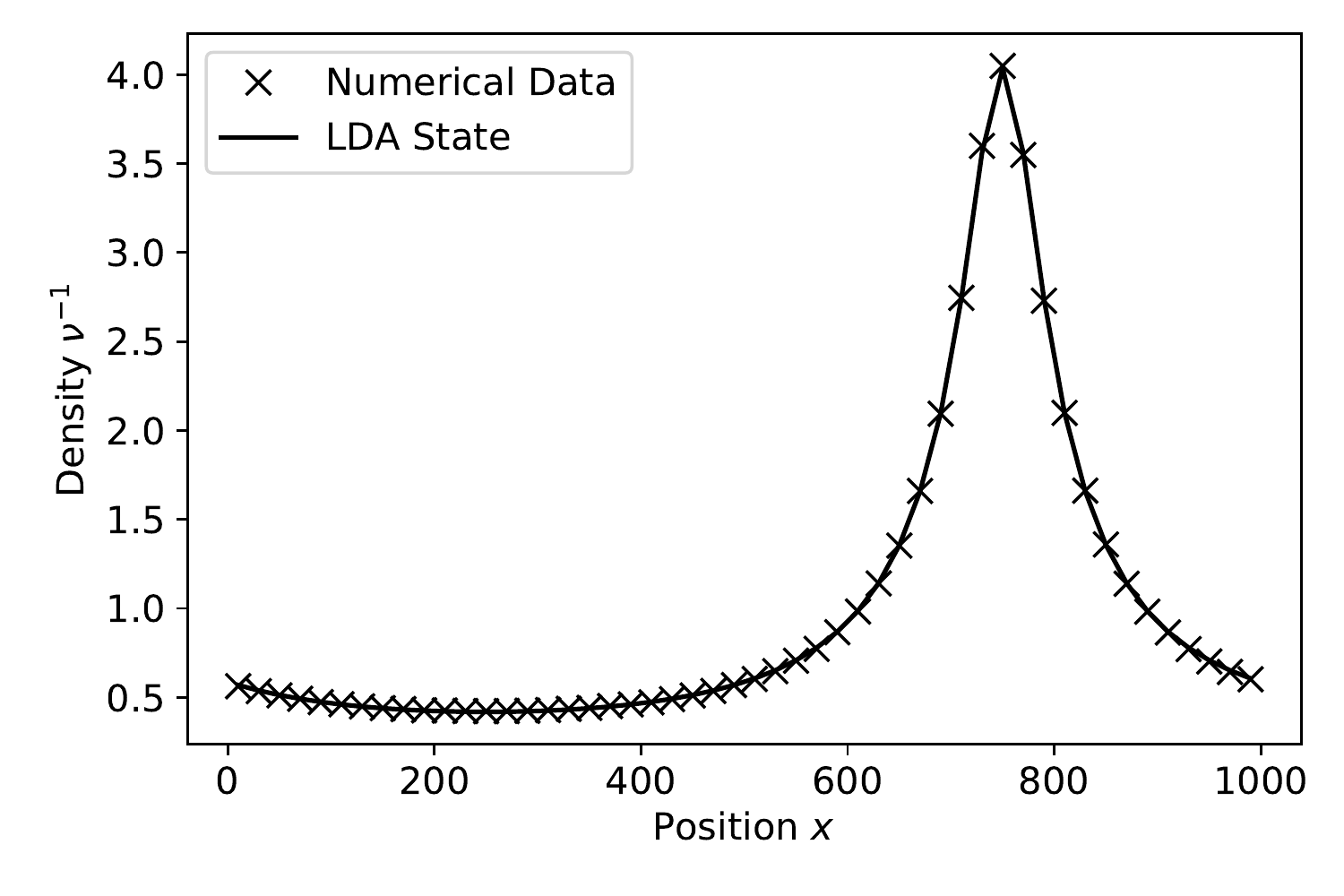}
    \includegraphics[width=0.485\textwidth]{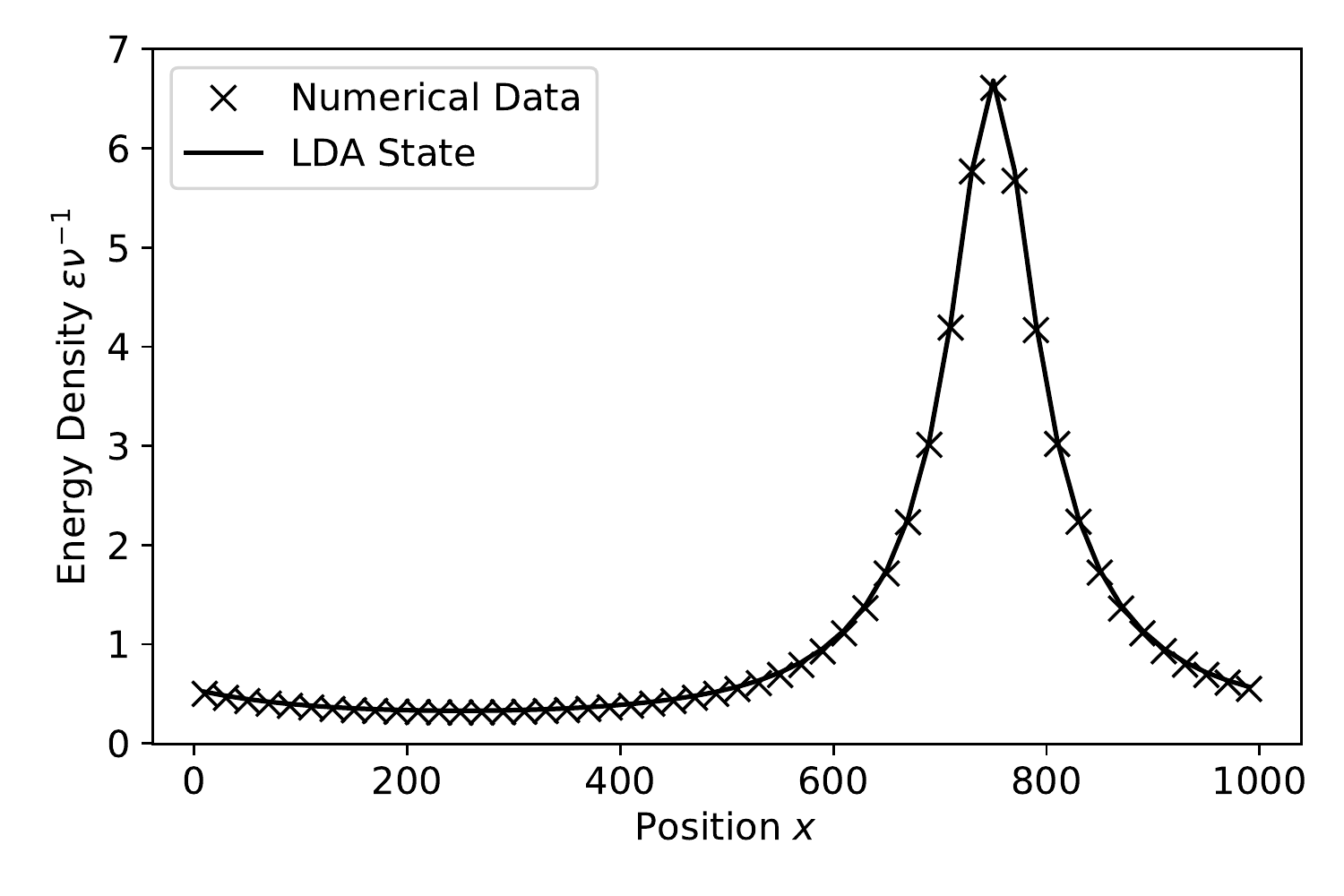}
    \caption{Numerical value of and LDA predictions for the energy density and number density of the stationary state. In this figure $N=1000$.}
    \label{fig:LDA}
\end{figure}
\begin{figure}
    \centering
    \includegraphics[width=0.75\textwidth]{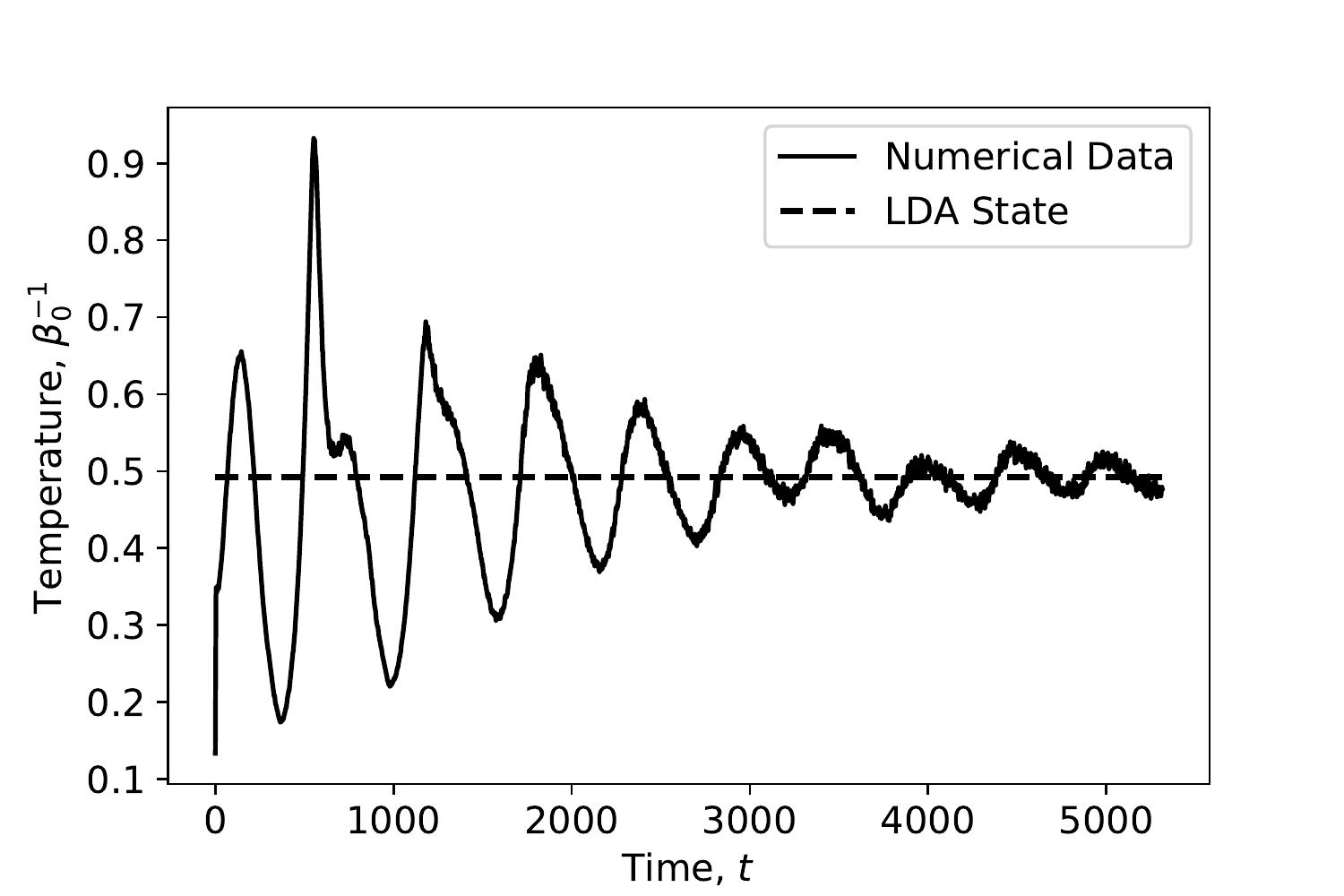}
    \caption{Convergence of the global temperature $\beta_0^{-1}=N^{-1}\sum_{i=1}^NV(x_i)p_i^2$, which is equal to the thermodynamic quantity by equipartition, to the LDA value. In this figure $N=1000$.}
    \label{fig:temp}
\end{figure}
\begin{figure}
    \centering
    \includegraphics[width=0.485\textwidth]{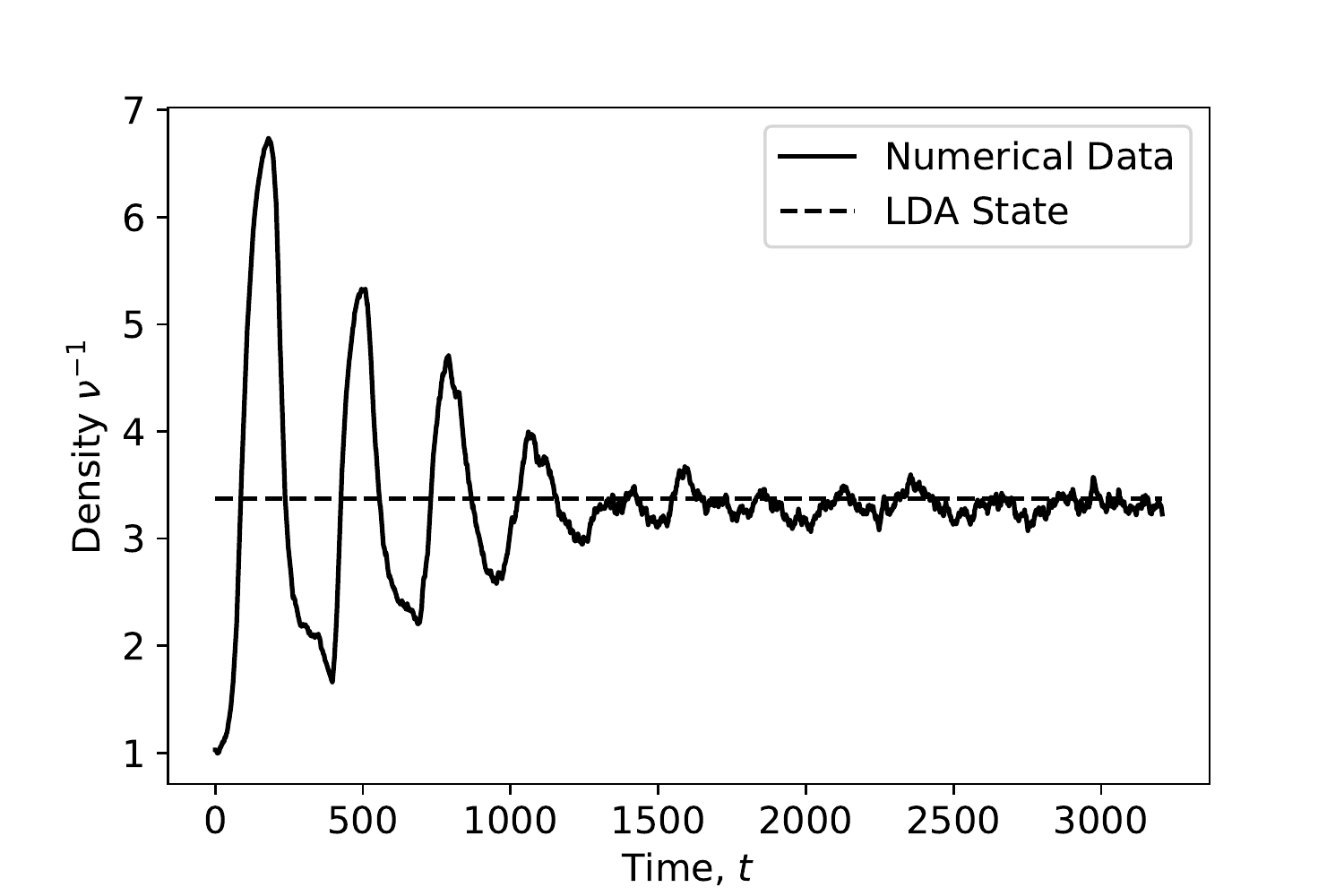}
    \includegraphics[width=0.485\textwidth]{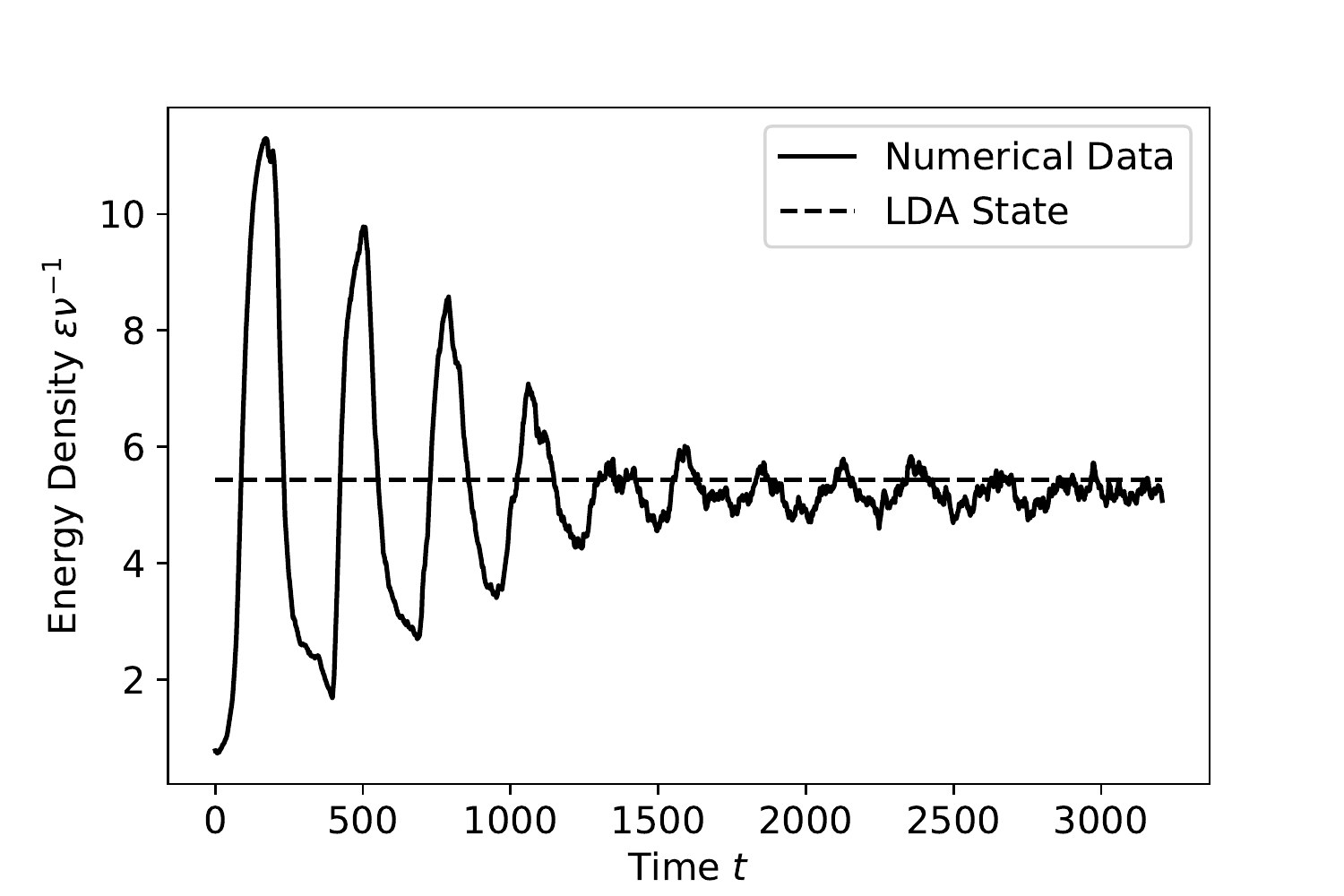}
    \caption{Convergence of the density and energy density near the trough of the potential to the LDA prediction. In this figure $N=500$.}
    \label{fig:density_convergence}
\end{figure}
\begin{figure}
    \centering
    \includegraphics[width=0.485\textwidth]{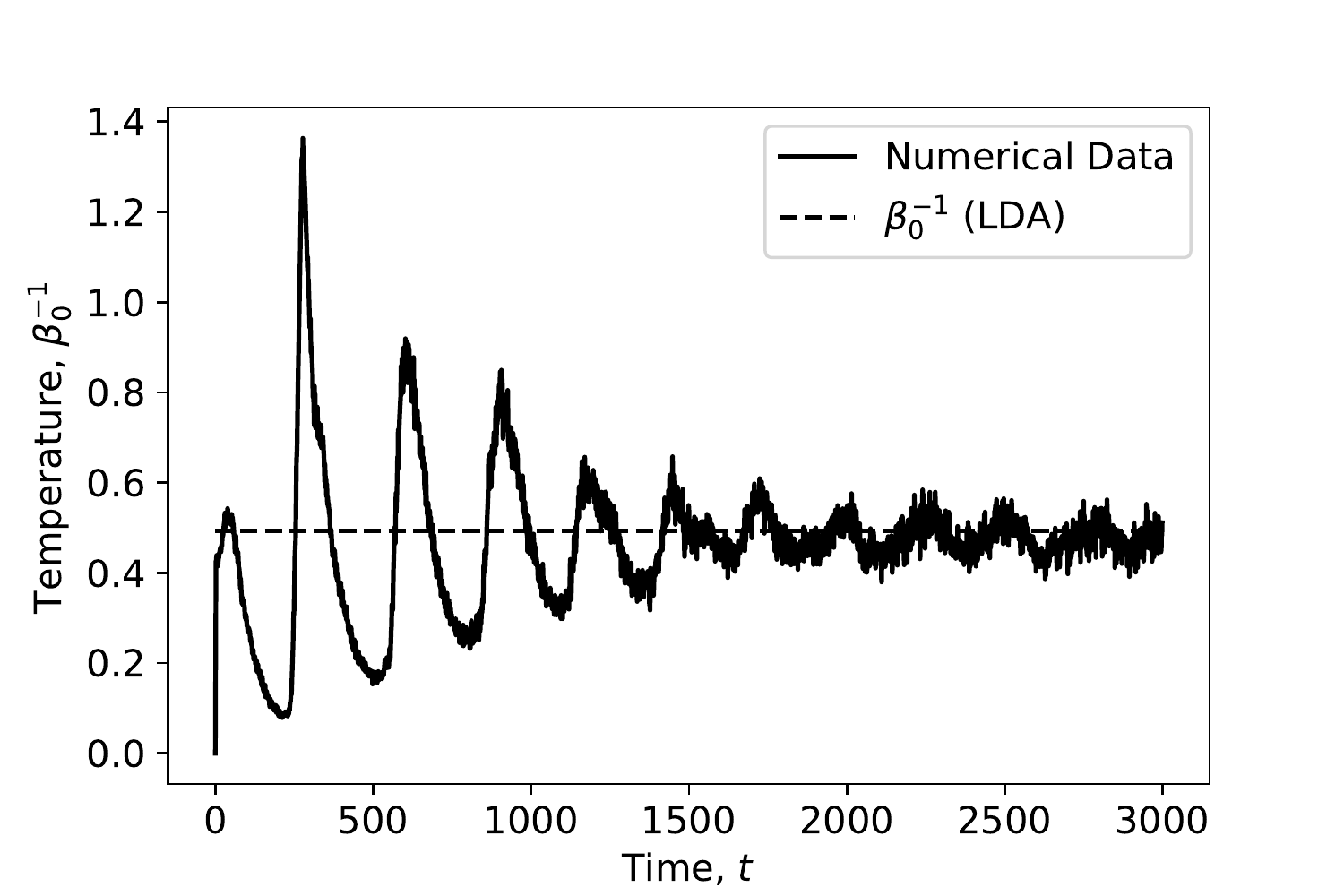}
    \includegraphics[width=0.485\textwidth]{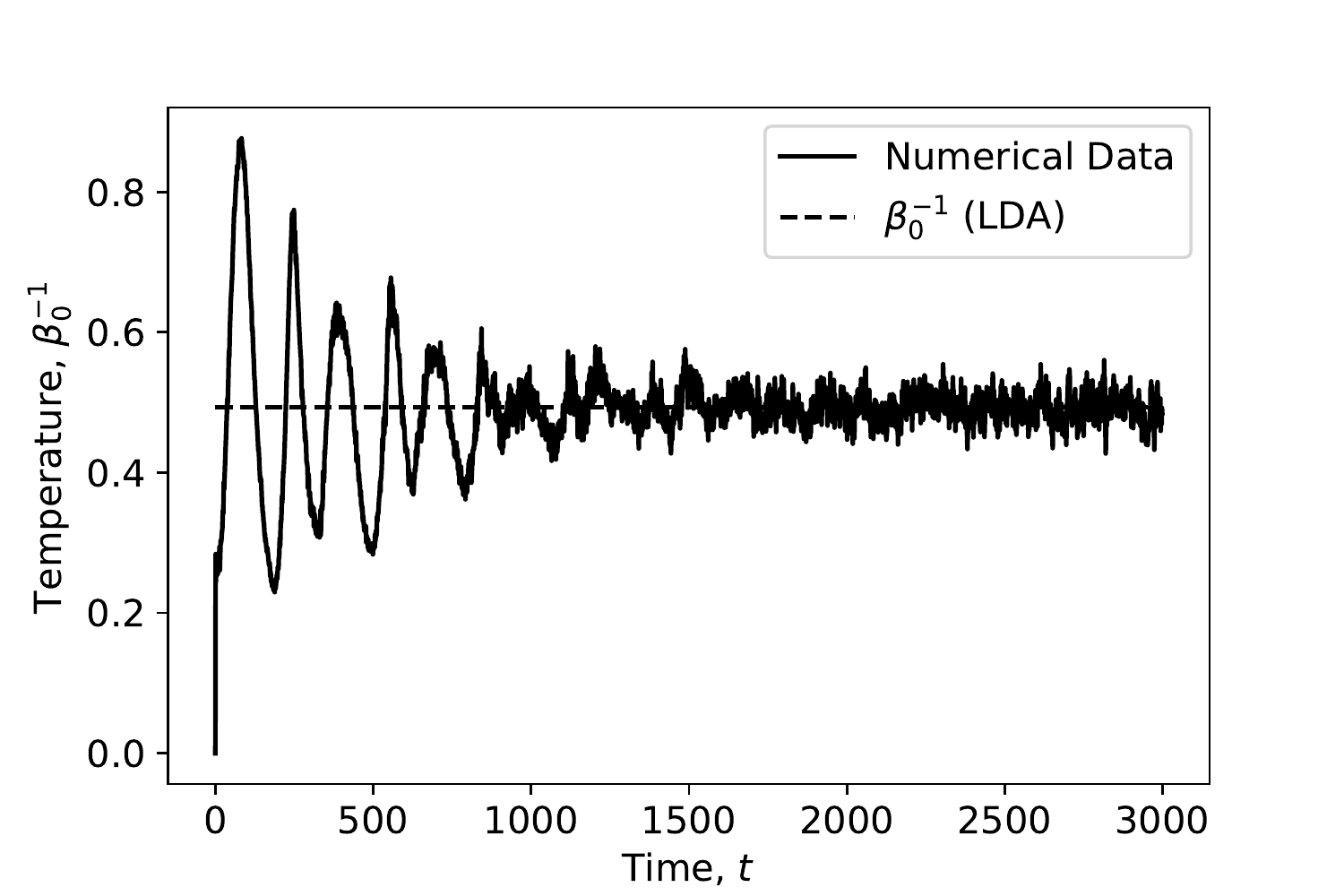}
    \caption{Position independence of the temperature $\beta_0^{-1}$, defined within a region as $\beta_0^{-1}=N^{-1}\sum_{i=1}^NV(x_i)p_i^2$. Shown is (L) the convergence to the LDA value for $0< x\le L/2$ and (R) for $L/2<x\le L$, with the same average values being obtained at late times in both regions. In this figure $N=500$.}
    \label{fig:temp_convergence}
\end{figure}
By implementing a molecular dynamics solver for the Toda system with Hamiltonian \eqref{Toda_Ham}, we can show that the thermal LDA state \eqref{LDA_state} is indeed approximately reached at diffusive timescales. We take the periodic potential $V(x)=2.5+\sin(2\pi x/L)$, where the periodicity of the potential is equal to the periodicity of the Toda gas. We take $N=L$ throughout, with $N=1000$ in Figs.~\ref{fig:LDA} and \ref{fig:temp}, and $N=500$ in Figs.~\ref{fig:density_convergence} and \ref{fig:temp_convergence}, where more realisations are needed to obtain relatively smooth plots\footnote{The required simulation length scales $\sim N^3$, $N$ with the number of particles, and a further factor $N^2$ associated to the simulation time required to access diffusive timescales, assuming the system is held at constant density.}. Finally we take 50 fluid cells in both cases. The energy density $\exv{H}/N$ and density $\nu^{-1}$ are constant across initial states, with values of $1.26$ and $1$ respectively. 

In all the figures there are periodic oscillations of observables. By running the simulations without the external energy field, that is, for the usual Toda gas, it was verified that this is the period related to the periodic boundary conditions. In Fig.~\ref{fig:temp_convergence}, the two graphs have distinct frequencies, around the trough of the potential the dominant period is $\tau\sim 300$, and around the peak of the potential the dominant period is, to a good approximation, $\tau/2$. Comparing the predictions of the LDA state with the averages of the numerical data, we see a good agreement, to less than $5\%$ errors, of all measured quantities.

We note finally that a more complete evaluation of the descriptive powers of our formalism is available. The equations of section \ref{sec:int_systems} can be applied to the Toda gas by taking the statistical factor to be $f(\theta)=1$, and the scattering shift as $T(\theta,\alpha)=2\log|\theta-\alpha|$. We leave this task for a future publication.

\section{Conclusion}
We have introduced a fully general formalism for the hydrodynamics of a Hamiltonian system with generic inhomogeneous force fields, including terms to second order in spatial derivatives. The resulting hydrodynamic equation is fully expressed in terms of Euler currents and extended Onsager coefficients, which provide the net entropy increase, shown to be always non-negative. While our hydrodynamic equation applies to any inhomogeneous deformed Hamiltonian constructed from PT-symmetric densities of conserved quantities in involution, in the particular case of integrable systems { we showed that expressions for the extended Onsager coefficients may be obtained by a form factor expansion, and we obtained explicit expressions at the two particle-hole order, which are valid at low density of excitations.} We have thereby extended the equations of generalised Hydrodynamics to include generic force fields and dissipative terms. The equation has thermal states as the only stationary states and positive entropy increase, showing how second order terms in the hydrodynamics, namely diffusive processes in real and quasi-momenta space are responsible for the thermalisation of the system. This confirms and generalises our previous result for the one dimensional Bose-gas in a generic trapping potential \cite{PhysRevLett.125.240604}. We have confirmed the final thermalisation of an integrable classical system, a Toda chain under an inhomogeneous energy field representing the effect of an inhomogeneous Hamiltonian. 

Several extension and open questions are in reach. First the lower bound for the diffusion constant $\mathfrak{D}_{\boldsymbol{\mathfrak{f}^2}}\mathsf C$ we found should be tested against numerical predictions to check its validity. It is reasonable to expect that its corrections are subleading at low density of excitations and in weakly interacting limits \cite{Durnin2020,10.21468/SciPostPhys.9.6.082}. In these regimes the final equation \eqref{main_particle_result} can be applied straightforwardly to describe the effect of external magnetic and electric fields in lattice systems as the XXZ spin-$1/2$ chain and the Fermi Hubbard chain, where interesting non-equilibrium phenomena can be observed \cite{2010.12965,2102.01675}.  Moreover it will be important to clarify the role of diffusion in quasi-momentum space for the quasi-particles. The recent years have brought up different classes of non-diffusive transport in real space for integrable or quasi-integrable spin chains, related to the KPZ universality class, see \cite{Scheie2021,PhysRevB.101.041411,PhysRevLett.122.210602,2009.08425,Gopalakrishnan2019,2103.01976}. Similar forms of super-diffusion could be found in the quasi-momentum $\theta$ space, in particular in those cases where the kernel $ \mathfrak{D}_{\boldsymbol{\mathfrak{f}^2}}$ can diverge. Finally, the techniques we have used are clearly generalisable to higher-order hydrodynamics, which is thought to be meaningful at least in integrable systems. We shall investigate these exciting questions in the near future.  

\subsection*{Acknowledgements}
We acknowledge relevant discussions with Romain Vasseur, Sarang Gopalakrishnan, Jerome Dubail, Herbert Spohn. JD acknowledges funding from the EPSRC Centre for Doctoral  Training in Cross-Disciplinary Approaches to Non-Equilibrium Systems (CANES) under grant EP/L015854/1.

\begin{appendix}

\section{Computation of the Onsager coefficients }\label{sec:FFComputation}
Onsager coefficients in integrable models can be analytically computed by expanding over intermediate quasiparticle excitations, written in terms of particle-hole intermediate states. This corresponds to hydrodynamic projections of each current on the space of normal modes and their quadratic fluctuations, which has been shown to provide generically a lower bound for the Onsager coefficients \cite{1912.01551}. However, whenever at least one of the two currents in the Onsager coefficient is generated by the the homogeneous Hamiltonian, $w^i Q_i$, flow, the lower bound is saturated and the two-particle hole contribution provides the full Onsager coefficient (while the one particle-hole contribution gives the subtracted ballistic  part \cite{RevCorrelations}). We shall here write the calculation for such contribution for generic currents  
\begin{align} \label{eq:fullOnsager}
 \mathfrak L [i,j;j_{k,
    l}] = \lim_{t \to \infty}   &   \int_{-t}^{t} ds     (     J_{i,j }(y,s \ww   ),     j_{k,l}(0,0) )^C  \nonumber = \sum_{n=2}^\infty \mathfrak L_{n-\rm ph}[i,j;j_{k,
    l}]  \\&   =  \frac{ (2\pi)^2}{2!^2}  \lim_{t \to \infty}   \int \dd \theta_1^- \dd  \theta_2^-  \rho_{\text{p}}(\theta_1^-)  \rho_{\text{p}}(\theta_2^-)    \fint \dd \theta_1^+ \dd  \theta_2^+ \rho_{\text{h}}(\theta_1^+)   \rho_{\text{h}}(\theta_2^+)    \no \\& \times \delta(k)  \delta_t(\varepsilon_{\boldsymbol{w}})   \langle \rho_{\rm p} | j_{i,j} |  \theta^+_1, \theta^+_2, \theta^-_1, \theta^-_2 \rangle \langle \theta^+_1, \theta^+_2, \theta^-_1, \theta^-_2  | j_{k,l} | \rho_{\text{p}}   \rangle + \ldots,
\end{align}
where the integration $\fint$ denotes Hadamard regularisation, see for example \cite{Panfil2021}, due to the singularities in the integrand. 
In particular we shall use the matrix elements of the generalised current operators 
\begin{align}
& \langle \rho_{\rm p}| j_{k,i } | \{ \theta_{\rm p}^\bullet, \theta_{\rm h}^\bullet  \}\rangle = h^{\rm Dr}_k(\{ \theta_{\rm p}^\bullet, \theta_{\rm h}^\bullet  \})  f_i(\{ \theta_{\rm p}^\bullet, \theta_{\rm h}^\bullet  \})  .
\end{align}
 The function $h^{\rm Dr}_k$ is the eigenvalue of the $k$-th charge dressed by the shift function on the background given by $\rho_{\text{p}}$, see for example \cite{10.21468/SciPostPhys.6.4.049}, and the function $f_i$ is known and given by \cite{10.21468/SciPostPhys.6.4.049}
\begin{align}
    f_i(\{ \theta_{\rm p}^\bullet, \theta_{\rm h}^\bullet  \}) &=	\left(\frac{T^{\rm dr}(\theta^-_2, \theta^-_1) h_i^{\rm dr}(\theta^-_2) }{k'(\theta^-_1) k'(\theta^-_2) (\theta^+_1 - \theta^-_1)}+ \frac{T^{\rm dr}(\theta^-_1, \theta^-_2) h_i^{\rm dr}(\theta^-_1) }{k'(\theta^-_2) k'(\theta^-_1) (\theta^+_2 - \theta^-_2)} \right. \nonumber\\
	&\left. + \frac{T^{\rm dr}(\theta^-_2, \theta^-_1) h_i^{\rm dr}(\theta^-_2) }{k'(\theta^-_1) k'(\theta^-_2) (\theta^+_2 - \theta^-_1)} + \frac{T^{\rm dr}(\theta^-_1, \theta^-_2) h_i^{\rm dr}(\theta^-_1) }{k'(\theta^-_2) k'(\theta^-_1) (\theta^+_1 - \theta^-_2)} + (\dots) \right),
\end{align}
with corrections $(\dots)$ given by finite elements in the limit of one of the particles $\theta_i^+$ approaching the value of one of the holes $\theta_j^-$.   
Notice that the energy constraint need particular care. We reguralise it as follows 
\begin{equation}
   \int_{-t}^t ds \, e^{i s \varepsilon  }   = \frac{\sin(t \varepsilon)}{\pi \varepsilon} = \delta_t (\varepsilon).
\end{equation}
We proceed analogously to the standard case. Due to the two delta functions, the integral is around particle and holes having the same values. We therefore expand over small $\Delta_i = \theta_i^+ - \theta_i^-$ (and its permutation, accounting for an additional factor 2 to the integrated correlator), obtaining 
\begin{equation}
  k =k'(\theta_1) \Delta_1 +k'(\theta_2) \Delta_2 + \ldots,
\end{equation}
\begin{equation}
    \varepsilon_{\boldsymbol{w}} = v_{\boldsymbol{w}}^{\rm eff}(\theta_1) k'(\theta_1) \Delta_1 + v_{\boldsymbol{w}}^{\rm eff}(\theta_2) k'(\theta_2) \Delta_2 + \ldots,
\end{equation}
\begin{equation}
    h^{\rm Dr}_k = (h_{k}')^{\rm dr}(\theta_1)  \Delta_1 + (h'_{k})^{\rm dr}(\theta_2)   \Delta_2 + \ldots.
\end{equation}
The integration over $\Delta_2$ can be done using $\delta(k)$, which set $\Delta_2 = - \Delta_1 k'(\theta_1)/k'(\theta_2)$, which gives 
\begin{align} 
 \mathfrak L_{2-\rm ph}[i,j;j_{k,
    l}]  \nonumber \\&   =  \frac{1}{2} \lim_{t \to \infty} \int d\theta_1 \int d \theta_2 \fint d\Delta_1 \nonumber \\&  k'(\theta_1) n(\theta_1 )  f(\theta_1+ \Delta_1) k'(\theta_2) n(\theta_2) f(\theta_2- k'(\theta_1)/k'(\theta_2) \Delta_1) (T^{\rm dr}(\theta_1,\theta_2))^2 \nonumber \\&  \times \delta_t (\Delta_1 k'(\theta_1) (v_{\boldsymbol{w}}^{\rm eff}(\theta_1)  - v_{\boldsymbol{w}}^{\rm eff}(\theta_2)) ) k'(\theta_1) {(v^{\rm eff}_i(\theta_1) - v^{\rm eff}_i(\theta_2)) (v^{\rm eff}_k(\theta_1) - v^{\rm eff}_k(\theta_2)) }{} \nonumber \\& \times \left( \frac{h_j^{\rm dr}(\theta_1)}{k'(\theta_1)} - \frac{h_j^{\rm dr}(\theta_2)}{k'(\theta_2)} \right)\left( \frac{h_l^{\rm dr}(\theta_1)}{k'(\theta_1)} - \frac{h_l^{\rm dr}(\theta_2)}{k'(\theta_2)} \right),
\end{align}
where $v^{\rm eff}_i(\theta) = (h_i' )^{\rm dr}(\theta)/k'(\theta)$.
The limit $t \to \infty$ can now be taken, excluding the zero-measure set of points where $(v_{\boldsymbol{w}}^{\rm eff}(\theta_1)  - v_{\boldsymbol{w}}^{\rm eff}(\theta_2))=0$. 
Then $\Delta_1$ can be integrated with the $\delta(\varepsilon_{\boldsymbol{w}})$, which produces the Jacobian factor $|k'(\theta_1) (v_{\boldsymbol{w}}^{\rm eff}(\theta_1)  - v_{\boldsymbol{w}}^{\rm eff}(\theta_2))|$. We then obtain 
\begin{align} 
  \mathfrak L_{2-\rm ph}[i,j;j_{k,
    l}]  \nonumber \\&   =  \frac{1}{2} \int d\theta_1 \int d \theta_2 k'(\theta_1) n(\theta_1)  f(\theta_1) k'(\theta_2) n(\theta_2) f(\theta_2) (T^{\rm dr}(\theta_1,\theta_2))^2\nonumber \\& \times \frac{(v^{\rm eff}_i(\theta_1) - v^{\rm eff}_i(\theta_2)) (v^{\rm eff}_k(\theta_1) - v^{\rm eff}_k(\theta_2)) }{|v^{\rm eff}_{\boldsymbol{w}}(\theta_1) - v^{\rm eff}_{\boldsymbol{w}}(\theta_2) |} \nonumber \\& \times \left( \frac{h_j^{\rm dr}(\theta_1)}{k'(\theta_1)} - \frac{h_j^{\rm dr}(\theta_2)}{k'(\theta_2)} \right)\left( \frac{h_l^{\rm dr}(\theta_1)}{k'(\theta_1)} - \frac{h_l^{\rm dr}(\theta_2)}{k'(\theta_2)} \right).
\end{align}
  To recover the results in the main text it should be used also
\begin{equation}
    \beta^i v^{\rm eff}_i(\theta) = - n'(\theta)/(f(\theta) n(\theta) k'(\theta)),
\end{equation}
and 
\begin{equation}
    w^i v^{\rm eff}_i(\theta) = v^{\rm eff}_{\boldsymbol{w}}(\theta).
\end{equation}
One may also write an expression of the two particle-hole contribution to the generalised Onsager coefficient, in terms of the kernel \eqref{eq:ffun}, in the explicitly multilinear form
\begin{align}\label{Labcd}
   &\mathfrak L_{2-\rm ph} [\boldsymbol a,\boldsymbol b;j_{\boldsymbol c,
    \boldsymbol d}]
   \nonumber \\& =
    \frc12 \int d\theta_1\int d \theta_2\,
    \kappa_{\mathfrak D}(\theta_1,\theta_2) 
    \frc{\Delta v^{\rm eff}_{\boldsymbol a}(\theta_1,\theta_2)
    \Delta v^{\rm eff}_{\boldsymbol c}(\theta_1,\theta_2)
    \Delta a^{\rm eff}_{\boldsymbol b}(\theta_1,\theta_2)
    \Delta a^{\rm eff}_{\boldsymbol d}(\theta_1,\theta_2)}{(\Delta v^{\rm eff}_{\boldsymbol w}(\theta_1,\theta_2))^2},
\end{align}
where
\beq
     \Delta v^{\rm eff}_{\boldsymbol a}(\theta_1,\theta_2) = v^{\rm eff}_{\boldsymbol a}(\theta_1) - v^{\rm eff}_{\boldsymbol a}(\theta_2),\quad
     \Delta a^{\rm eff}_{\boldsymbol a}(\theta_1,\theta_2) = a^{\rm eff}_{\boldsymbol a}(\theta_1) - a^{\rm eff}_{\boldsymbol a}(\theta_2).
\eeq
By appropriately separating the integral kernel from the external functions, one immediately obtains \eqref{Lintegrable1}-\eqref{Lintegrable4}. Higher particle-hole contribution $\Big[ \mathfrak L[\boldsymbol a,\boldsymbol b;j_{\boldsymbol c,
    \boldsymbol d}] \Big]_{n>2-\rm ph} $ vanish whenever at least one of the form-factor is proportional to the Hamiltonian energy $\varepsilon_{\boldsymbol{w}}$, namely for current induced by Hamiltonian flow, due to the energy conservation $\delta(\varepsilon_{\boldsymbol{w}})$, which gives a finite contribution only for the the two-particle hole contribution, together with momentum conservation. However the case $\boldsymbol{a} = \boldsymbol{\beta}$ and $\boldsymbol{c} = i$  evades this case. Therefore the full sum over generic number of particle-hole excitations in this case remains 
    \begin{equation}
      \mathfrak L[\boldsymbol{ \beta},\boldsymbol {\mathfrak{f}};j_{i,
    \boldsymbol{ \mathfrak{f}}}]   =  \mathfrak L_{2-\rm ph}[\boldsymbol{ \beta},\boldsymbol {\mathfrak{f}};j_{i,
    \boldsymbol{ \mathfrak{f}}}] + \sum_{n \geq 3} \mathfrak L_{n-\rm ph}[\boldsymbol{ \beta},\boldsymbol {\mathfrak{f}};j_{i,
    \boldsymbol{ \mathfrak{f}}}]  .
    \end{equation}
As the sum over $n>2$ particle-hole contribution is currently out-of-reach, we shall here only provide the $n=2$ contribution, which constitutes a finite lower bound to the Onsager coefficient $\mathfrak L[\boldsymbol{ \beta},\boldsymbol {\mathfrak{f}};j_{i,
    \boldsymbol{ \mathfrak{f}}}]$. Such higher particle-hole contribution will not affect total energy conservation, see eq. \eqref{eq:energyconsconstraint} and the positiviness of the entropy increase.

\section{PT symmetry and perturbation theory}\label{app:PT}
To include all terms up to second derivative order in the hydrodynamic equation in the presence of inhomogeneous external fields, we must in principle also include the following term in the Hamiltonian \eqref{eq:pert}
\begin{equation}
    H^{(2)}=-\frac{\p_{x_0} \mathfrak{f}^i(x_0)}{2}\int dx\,(x-x_0)^2 \ q_i(x).
\end{equation}
However, in systems with PT symmetry the contribution to the hydrodynamics from this term vanishes. Taking expectation values with respect to the local homogeneous, PT invariant states (neither perturbation theory nor inhomogeneity of the state need be considered as we are already at highest order in derivatives), we find
\begin{align}
    \exv{[H^{(2)},q_i(x_0)]}=&-\frac{\p_{x_0} \mathfrak{f}^k(x_0)}{2}\int dx\,(x-x_0)^2\exv{[q_k(x),q_i(x_0)]}\nonumber \\
    =&\frac{\p_{x_0} \mathfrak{f}^k(x_0)}{2}\int dx\,x^2\exv{[q_k(x),q_i]}^*
    \nonumber \\
    =&\frac{\p_{x_0} \mathfrak{f}^k(x_0)}{2}\int dx\,x^2\exv{\mathcal{PT}([q_k(x),q_i])}
    \nonumber \\
    =&\frac{\p_{x_0} \mathfrak{f}^k(x_0)}{2}\int dx\,x^2\exv{[q_k(-x),q_i]}\nonumber \\
    =&\frac{\p_{x_0} \mathfrak{f}^k(x_0)}{2}\int dx\,x^2\exv{[q_k(x),q_i]},
\end{align}
which implies 
\begin{equation}
    \exv{[H^{(2)},q_i(x_0)]}=0.
\end{equation}
We therefore conclude that terms proportional to $\p_{x_0} \mathfrak{f}^i(x_0)$ only can appear at $3$-th order in spatial derivatives, within the hydrodynamic gradient expansion.

\end{appendix}

\bibliographystyle{ieeetr.bst}
\bibliography{biblio}

\end{document}